\documentclass[useAMS,usenatbib]{mn2e}
\usepackage{graphicx}
\DeclareGraphicsExtensions{.pdf}
\usepackage{amsfonts,amsmath,amssymb}

\usepackage[colorlinks=true,linkcolor=red, citecolor= blue, urlcolor=cyan]{hyperref}
 
\usepackage{upgreek}

\newcommand {\aplt} {\ {\raise-.5ex\hbox{$\buildrel<\over\sim$}}\ } 
\newcommand{\be}{\begin{equation}}
\newcommand{\ee}{\end{equation}}

\newcommand{\mincir}{\raise
  -2.truept\hbox{\rlap{\hbox{$\sim$}}\raise5.truept \hbox{$<$}\ }}
\newcommand{\magcir}{\raise
  -2.truept\hbox{\rlap{\hbox{$\sim$}}\raise5.truept \hbox{$>$}\ }}
\newcommand{\siml}{\raise
  -2.truept\hbox{\rlap{\hbox{$\sim$}}\raise5.truept \hbox{$<$}\ }}
\newcommand{\simg}{\raise
  -2.truept\hbox{\rlap{\hbox{$\sim$}}\raise5.truept \hbox{$>$}\ }}

\begin{document}

\title[Calibration of CODEX Richness-Mass relation]{Mass Calibration of the CODEX Cluster Sample using SPIDERS Spectroscopy - I. The Richness-Mass Relation}

\newcommand{\Munich}{$^{1}$}
\newcommand{\ExcellenceCluster}{$^{2}$}
\newcommand{\MPE}{$^{3}$}
\newcommand{\Trieste}{$^{4}$}
\newcommand{\IRAP}{$^{5}$} 
\newcommand{\Helsinki}{$^{6}$} 
\newcommand{\Liverpool}{$^{7}$}

\author[Capasso et al.] {R.~Capasso\thanks{Raffaella.Capasso@physik.lmu.de}\Munich$^,$\ExcellenceCluster,
J.~J.~Mohr\Munich$^,$\ExcellenceCluster$^,$\MPE,
A.~Saro\Munich$^,$\ExcellenceCluster$^,$\Trieste,
A.~Biviano\Trieste,
N.~Clerc\MPE$^,$\IRAP,
\newauthor
A.~Finoguenov\MPE$^,$\Helsinki,
S.~Grandis\Munich$^,$\ExcellenceCluster,
C.~Collins\Liverpool,
G.~Erfanianfar\MPE,
S.~Damsted\Helsinki,
\newauthor
C.~Kirkpatrick\Helsinki,
A.~Kukkola\Helsinki
\\
\\
\Munich Faculty of Physics, Ludwig-Maximilians-Universit\"{a}t, Scheinerstr.\ 1, 81679 Munich, Germany \\
\ExcellenceCluster Excellence Cluster Universe, Boltzmannstr.\ 2, 85748 Garching, Germany \\
\MPE  Max Planck Institute for Extraterrestrial Physics, Giessenbachstr.\ 85748 Garching, Germany \\
\Trieste INAF-Osservatorio Astronomico di Trieste via G.B. Tiepolo 11, 34143 Trieste, Italy \\
\IRAP IRAP, Université de Toulouse, CNRS, UPS, CNES, Toulouse, France \\
\Helsinki Department of Physics, University of Helsinki, Gustaf H\"{a}llstr\"{o}min katu 2a, FI-00014 Helsinki, Finland \\
\Liverpool Astrophysics Research Institute, Liverpool John Moores University, IC2, Liverpool Science Park, 146 Brownlow Hill, Liverpool, L3 5RF, UK
}

\pubyear{2018}
\date{\today}
\maketitle               
\label{firstpage}

\begin{abstract} 
We use galaxy  dynamical information to calibrate the richness--mass scaling relation of a sample of 
428 galaxy clusters that are members of the CODEX sample with redshifts up to $z\sim0.7$.  
These clusters were X-ray selected using the ROSAT All-Sky Survey (RASS) and then cross-matched 
to associated systems in the redMaPPer catalog from the Sloan Digital Sky Survey.
The spectroscopic sample we analyze was obtained in the SPIDERS program and
contains $\sim$7,800 red member galaxies.  Adopting NFW mass and galaxy density profiles and a 
broad range of orbital anisotropy profiles, we use the Jeans equation to calculate halo masses. 
Modeling the scaling relation as $\lambda \propto \text{A}_{\lambda}  {M_{\text{200c}}}^{\text{B}_{\lambda}}  ({1+z})^{\gamma_{\lambda}}$, 
we find the parameter constraints 
$\text{A}_{\lambda}=38.6^{+3.1}_{-4.1}\pm3.9$, $\text{B}_{\lambda}=0.99^{+0.06}_{-0.07}\pm0.04$, 
and $\gamma_{\lambda}=-1.13^{+0.32}_{-0.34}\pm0.49$, where we present systematic uncertainties 
as a second component. 
We find good agreement with previously published mass trends with the exception of those from
stacked weak lensing analyses.  We note that although the lensing analyses failed to account for the 
Eddington bias, this is not enough to explain the differences.  We suggest that
differences in the levels of contamination between 
pure redMaPPer and RASS+redMaPPer samples could well contribute to these differences.  The redshift trend 
we measure is more negative than but statistically consistent with previous results.  We suggest that 
our measured redshift trend reflects a change in the cluster galaxy red sequence fraction with redshift, 
noting that the trend we measure is consistent with but somewhat stronger than an independently 
measured redshift trend in the red sequence fraction.  We also examine the impact of a plausible 
model of correlated scatter in X-ray luminosity and optical richness, showing it has negligible impact 
on our results.
\end{abstract}



\section{Introduction}  
\label{sec:introduction}

The formation and evolution of galaxy clusters is governed by the complex interplay between the gravity-induced dynamics of collapse and the baryonic processes associated with galaxy formation. Galaxy clusters, thus, constitute unique laboratories for both astrophysics and cosmology. On one side, the abundance of these objects as a function of mass and redshift is a well established cosmological probe \citep[e.g.,][]{1993White,2001Haiman,2015Mantz,2016deHaan}. On the other side, the observation of the evolution of galaxy properties in clusters provide us with information on galaxy formation, their assembly history, and the correlation between their evolution and environment \citep[e.g.,][]{1984Dressler,1999Depropris,2009Mei, 2012Muzzin,2017Hennig,2018Strazzullo,2019Capasso}. 

Of primary importance to both types of studies are accurate mass estimates and large samples of clusters with well understood selection.  For cosmological studies that adopt the halo mass function this is obvious, but for galaxy population studies it is equally important, because galaxy properties vary with clustercentric distance, and thus to compare properties of clusters across a range of mass and redshift, it is crucial to be able to adopt a meaningful overdensity radius such as $r_\mathrm{200c}$, which 
corresponds to the radius at which the mean enclosed density is 200 times the critical density and is thus trivially derived from the corresponding mass $M_\mathrm{200c}$.  Adopting an overdensity radius reveals cluster regularity or approximate self-similarity in structure formation simulations \citep[e.g,][]{1997Navarro} and has also revealed regularity in studies of real clusters \citep[e.g.][]{2007Pratt}.

A good understanding of the mass--observable relation that links the mass of galaxy clusters to readily obtainable observables such as the optical richness $\lambda$ is then more than a convenience.  It enables both cosmological and structure formation studies on large cluster ensembles. Within this context, uncertainties on cluster masses include the measurement uncertainties on the observable, the intrinsic scatter in the observable at fixed mass and redshift and the uncertainties on the parameters of the mass--observable relation.  The latter can be controlled through calibration. 

Different mass constraints have been used to calibrate the mass--observable relation for cluster ensembles, each with its advantages and disadvantages. 
Weak lensing distortions of background galaxies by clusters can be used to provide accurate cluster mass estimates  \citep[e.g.][]{2009Corless, 2011Becker, 2018Dietrich, 2018McClintock}. However, mass measurements from weak gravitational lensing of background galaxies become extremely challenging at high redshift $z\sim1$, where the number of background sources in typical imaging datasets drops, weakening the mass constraints.  Moreover, the scatter between weak lensing inferred masses and true halo mass is large, implying that large numbers of clusters are needed for accurate mass calibration. Recently, \citet{2018Baxter} applied gravitational lensing of the Cosmic Microwave Background (CMB), using CMB maps from the South Pole Telescope (SPT) 2,500~deg$^2$ SPT-SZ survey, demonstrating an ability to constrain the amplitude of the $\lambda$--mass relation to $\sim$20\% accuracy.  This offers great promise for the future, assuming systematic biases due to the thermal Sunyaev-Zel'dovich effect and cluster mis-centering can be accurately corrected. Cluster velocity dispersions, obtained through spectroscopic observations of cluster member galaxies, have proven to be good mass proxies as well, due in part to their insensitivity to complex ICM physics. But as with weak lensing masses, dispersion based masses still show large per-cluster scatter \citep{2008Evrard, 2013Saro, Sifon2013, 2014Ruel}, implying that large samples must be used for mass calibration.

In this work, we aim to calibrate the $\lambda$--mass--redshift scaling relation parameters by performing a dynamical analysis based on the Jeans equation \citep{1987Binney}. In particular, we use a modification of the MAMPOSSt  technique \citep[Modeling Anisotropy and Mass Profiles of Observed Spherical Systems;][]{2013MAMPOSSt}, which fits the distribution of particles in the observed projected phase space (line of sight velocities and distribution as a function of projected radius), to use the full information in the LOS velocity distribution and projected positions of cluster galaxies.  This method has been extensively used to recover dynamical masses and gain information on galaxy formation and evolution \citep[e.g.][]{2013Biviano,2017Biviano, 2014Munari}. In particular, in \citet{2019Capasso} it was demonstrated that, using this method on a composite cluster with $\sim$600 cluster members, dynamical masses and orbital anisotropy of the galaxy population can be simultaneously constrained, delivering masses with a $\sim$15\% uncertainty (decreasing to $\sim$ 8\% when using a composite cluster with $\sim$ 3000 tracers).  In addition, it was shown that combining cluster dynamical constraints in likelihood space produces final mass constraints that are consistent with masses from composite or stacked cluster analyses.

We perform a dynamical analysis on the ROSAT All-Sky Survey (RASS) X-ray cluster candidates, which have optical counterparts in SDSS imaging data identified using the redMaPPer algorithm \citep[the red sequence Matched-filter Probabilistic Percolation algorithm,][see Section~ \ref{sec:RM}]{2014Rykoff}.
The resulting cluster catalog is called CODEX (COnstrain Dark Energy with X-ray clusters; Finoguenov, in prep), and a subset of these clusters have since been spectroscopically 
studied within the SPectroscopic IDentification of eRosita Sources (SPIDERS) survey \citep{2016Clerc}. The analysis carried out here includes a sample of 428 CODEX clusters with a corresponding sample of $\sim$7,800 red member galaxies with measured redshifts.  The clusters span the redshift range $0.03 \leq z_{\text{c}}  \leq 0.66$, with richness $ 20 \leq \lambda \leq 230$. 

The paper is organized as follows: In Section~\ref{sec:data} we summarize the dataset used for our analysis. In Section~\ref{sec:theory} we give an overview of the theoretical framework. The results are presented in Section~\ref{sec:results}, where we discuss the outcome of our mass--observable relation calibration, and we present our conclusions in Section~\ref{sec:conclusions}.  Throughout this paper we adopt a flat $\Lambda$CDM cosmology with a Hubble constant $H_{0} = 70 \,  \text{km} \, \text{s}^{-1} \,  \text{Mpc}^{-1}$, and a matter density parameter $\Omega_{\text{M}} = 0.3$. Cluster masses ($M_{\text{200c}}$) are defined within 
$r_\mathrm{200c}$, the radius within which the cluster overdensity is 200 times the critical density of the Universe at the cluster redshift. We refer to $r_\mathrm{200c}$ simply as the virial radius. All quoted uncertainties are equivalent to Gaussian $1\sigma$  confidence regions unless otherwise stated.


\section{Data}
\label{sec:data}

This work is based on a spectroscopic galaxy sample constructed within the SPIDERS survey \citep{2016Clerc}, which observed a subset of CODEX galaxy clusters.  These clusters were selected from the ROSAT All-Sky Survey \citep[RASS, see][]{1999Voges} and then cross--matched with nearby optically selected systems identified using the redMaPPer algorithm applied to the Sloan Digital Sky Survey IV \citep[SDSS-IV, see][]{2016Dawson, 2017Blanton} optical imaging data.  In the following section we describe each of these elements of the dataset.

\subsection{The redMaPPer algorithm}
\label{sec:RM}

redMaPPer is an optical cluster-finding algorithm based on the red sequence technique, built around the richness estimator of \citet{2012Rykoff}. It has been successfully applied to photometric data from the Eighth Data Release \citep[DR8; ][]{2011Aihara} of the Sloan Digital Sky Survey, and subsequently to the SDSS Stripe 82 coadd data \citep{2014Annis} and to the Dark Energy Survey (DES) Science Verification Data (SV) and Year 1 (Y1) data \citep{2015Saro, 2016Rykoff, 2016Soergel}. It has been shown to provide excellent photometric redshift performance and optical richness estimates $\lambda$ that tightly correlate with external mass proxies.

The optical catalog construction is performed in several steps. First of all, the red sequence model is calibrated on a set of clusters having spectroscopic redshifts. This model is then used to identify galaxy clusters and measure their richness. To each galaxy in the vicinity of a galaxy cluster, redMaPPer estimates the membership probability, $\text{P}_{\text{mem}} \in$ [0, 1], based on its magnitude, colors and clustercentric distance. This probability is also used to estimate the richness of the cluster.  The 
latter is thus defined as the sum of the membership probabilities ($\text{P}_{\text{mem}}$) over all galaxies $\lambda = \sum \text{P}_{\text{mem}}$.

\subsection{The CODEX sample}
\label{sec:CODEX}

The CODEX survey is designed to combine ROSAT X-ray cluster candidates with optical selected cluster candidates identified using redMaPPer \citep[the red sequence Matched-filter Probabilistic Percolation algorithm,][see Section~ \ref{sec:RM}]{2014Rykoff}.   This catalog is constructed in several steps.  As a first step, RASS data are used to identify all X-ray sources with detection significance S/N$>4$.  The redMaPPer algorithm is then run on the SDSS imaging data around each RASS source position to identify candidate clusters with a red sequence, which constitutes a collection of passive galaxies lying at a common redshift.  The redMaPPer algorithm provides an estimate for the photometric redshift of the cluster, an estimation of the optical richness and an optical cluster center, which is constrained to be within {3\arcmin} of the X-ray position.  In cases of multiple optical counterparts meeting these criteria, the counterpart having the highest richness is assigned to the RASS X-ray source.  

Using the updated optical position of the cluster, a revised red sequence is identified, providing the final estimate of the cluster photometric redshift and richness (optical or ``OPT" quantities: $z_{\lambda,\text{OPT}}, \lambda_{\text{OPT}}$, etc.). If the cluster is at sufficiently high redshift that the SDSS photometric data are not deep enough to allow a direct measurement of richness over a fixed fraction of the cluster galaxy luminosity function (i.e., to a limit $m_*(z)+\Delta$, where $\Delta$ is the same for all clusters), then a correction factor $\eta$ is calculated and applied to the richness.  As described in Section~\ref{sec:scalingrelation}, this has an impact on the Poisson noise contribution to the richness and must be included in the analysis of the mass--observable scaling relation.

In the final step, X-ray properties based on the RASS count-rate and the redMaPPer redshift are calculated in optimized apertures (imposing a minimal signal-to-noise threshold of 1.6), assuming a  model for the X-ray spectral emissivity, along with the aperture-corrected cluster flux $f_{X}$ and  [0.1-2.4]~keV luminosities $L_{X}$. The final CODEX sample then results in X-ray detected clusters, for which we have an estimate of the redshift, optical richness, the optical cluster center, and X-ray luminosity. This sample has been used for follow-up observations of the SPIDERS survey, described below, which finally provided spectroscopic redshift estimates of cluster member galaxies.

\subsection{The SPIDERS spectroscopic sample}
\label{sec:spiders}

The SPIDERS survey is an observational program aiming to obtain homogeneous and complete spectroscopic follow-up of extragalactic sources, using data from X-ray satellites that lie within the SDSS-IV imaging footprint. The driving goals of the program are the confirmation of X-ray extended sources identified as galaxy cluster candidates and the assignment of a precise 
redshift. In the final years of SDSS-IV, SPIDERS will follow-up X-ray extended sources extracted from the all sky X-ray eROSITA survey \citep[extended ROentgen Survey with an Imaging Telescope Array][]{2010Predehl, 2012Merloni}. Prior to the launch of eROSITA, galaxy clusters identified in the shallower RASS and sparser XMM-Newton data will constitute the bulk of the SPIDERS program. The spectroscopy is obtained using the BOSS spectrograph mounted on the SDSS-2.5m telescope at Apache Point Observatory \citep{2006Gunn}, performing follow-up of galaxies detected in the large area of extragalactic sky imaged in \textit{ugriz} filters by the same telescope. In the following sections we describe the target selection, the cuts made on the sample, and how the spectroscopic galaxy sample used in this work is obtained. 

\subsubsection{Target selection}
\label{sec:targetselection}

The target selection and initial cuts to the sample are outlined in \citet{2016Clerc}. Here, we summarize the most salient features. To optimize the number of spectroscopically confirmed clusters, the redMaPPer membership probability $\text{P}_{\text{mem}}$ is used as a reference to assign priorities to potential targets, ranking galaxies within each cluster.  The algorithm starts with the richest cluster in the sample, iteratively proceeding to lower richness. The pool of targets along with the priority flag is then submitted to the eBOSS tiling algorithm. The final data reduction and spectral classification relies on the eBOSS spectroscopic pipeline and processing.

An automatic procedure is used to assign the membership of red sequence galaxies with measured redshifts. For each cluster, an iterative clipping procedure is performed. As a first step, members with velocity offsets greater than 5,000~km/s (relative to this first guess mean redshift) are rejected. The remaining potential members $N_{\text{z-spec}}$  are used to estimate the velocity dispersion of the cluster, either using the bi-weight variance \citep[$N_{\text{z-spec}} \geq$ 15; see][]{1990Beers} or the gapper estimator (if $N_{\text{z-spec}} < $  15). Finally, a $3\sigma$ clipping is applied, rejecting objects lying further away than 3 times the velocity dispersion from the mean velocity. 


A final validation of all galaxy clusters and assessment of their redshifts is achieved through visual screening of the outcome of the automatic procedure. Sometimes the automated procedure fails. This occures, for example,  if fewer than 3 members are assigned to a cluster, or if the initial 5,000~km/s clipping rejected all members. The latter can occur when there are several distinct structures along the line of sight.  Independent inspectors analyze these complex cases, which may lead to inclusion or removal of members.
This process sets the validation status and mean redshift of the cluster. Line-of-sight projection effects not disentangled by the photometric membership algorithm can also be identified and  split into several components. Final cluster redshift estimates are based on the bi-weight average \citep{1990Beers} of all red sequence galaxies selected as cluster members, if at least 3 members are assigned to the cluster. 
The typical cluster redshift statistical uncertainty is $\Delta_{z}/(1 + z) \lesssim 10^{-3}$.

The updated cluster spectroscopic redshifts are then used to update the computation of X-ray cluster properties. Assuming the standard flat $\Lambda$CDM cosmological model (Hubble constant $H_{0} = 70 \,  \text{km} \, \text{s}^{-1} \,  \text{Mpc}^{-1}$, and matter density parameter $\Omega_{\text{M}} = 0.3$, ROSAT fluxes are converted into rest-frame [0.1-2.4]~keV luminosities and scaling relations allow an estimate of the cluster mass and characteristic radius $r_\mathrm{500}$ or $r_\mathrm{200c}$. The typical measurement uncertainty on the luminosities of CODEX clusters amounts to $\approx$ 35\%, as computed from the Poissonian fluctuation in the associated ROSAT X-ray photons \citep[see][]{2015Mirkazemi}. 

\subsubsection{Final spectroscopic cluster member sample}
\label{sec:finalmemberselection}

Given the sample produced as described above, we apply some additional cuts prior to our analysis. As mentioned above, there are cases in which a CODEX cluster has multiple groups of galaxies separated by a large velocity gap along the line of sight. To avoid merging systems, we only use clusters which are flagged as having one component along the line of sight.  We restrict our analysis to the cluster virial region ($R  \leq r_\mathrm{200c}$). Moreover, we exclude the very central cluster region ($R \leq 50 \text{kpc}$), to account for the positional uncertainties of cluster centers, and to avoid including the centrally located BCG in the dynamical analysis.
At the end of this process, our spectroscopic dataset from SPIDERS consists of 705 galaxy clusters, for a total of $\approx 11,400$ candidate cluster members, with a median redshift $z = 0.21$ and spanning a richness range $20 \leq \lambda \leq 230 $. 
At the time this paper is being written, the observations of the galaxy clusters included in our sample have already been completed. No further galaxy spectroscopic redshifts will be assigned to these clusters during the final stages of the SDSS-IV program.

\begin{figure}
\centering 
\includegraphics[scale=0.50]{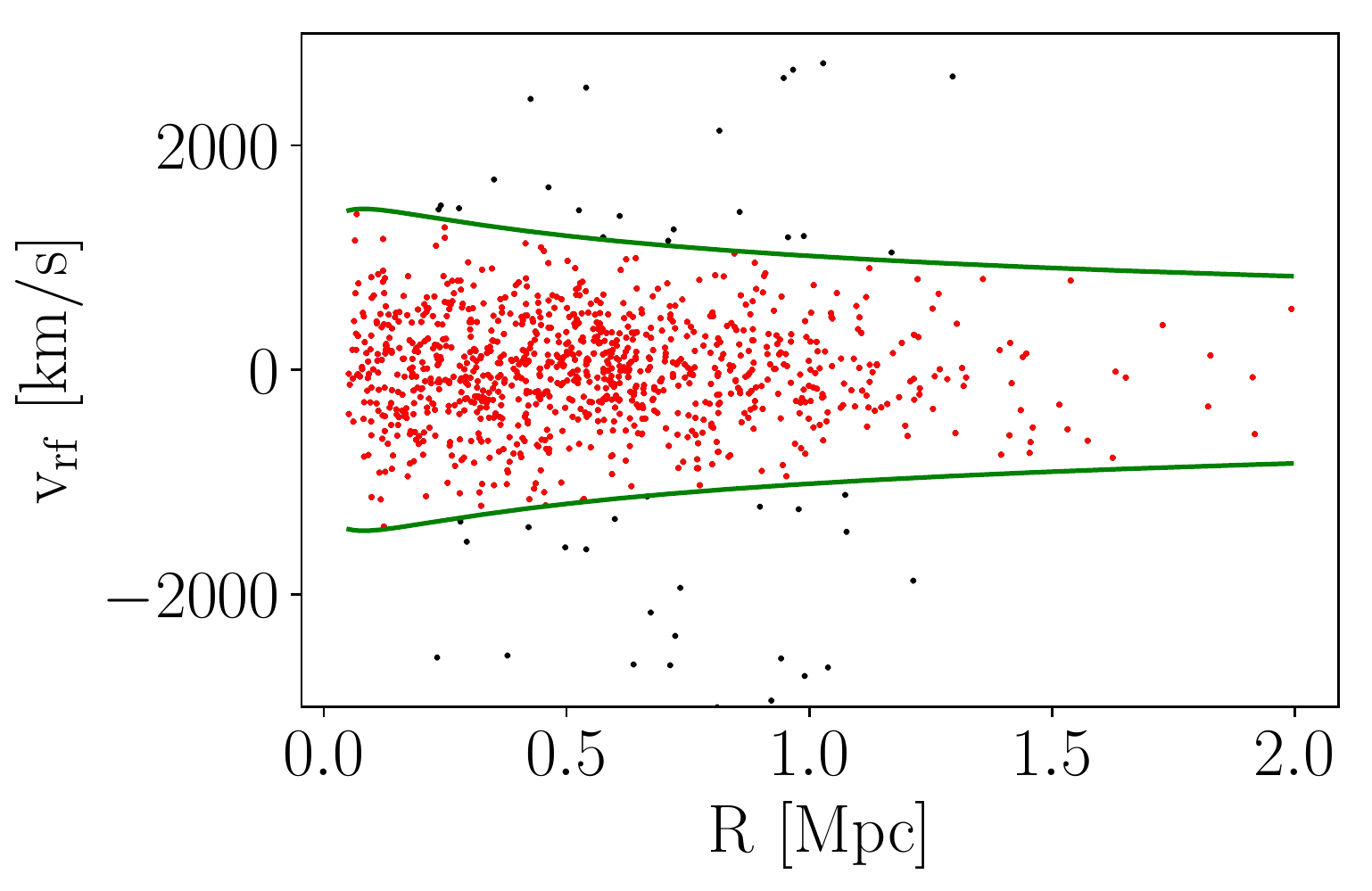}
\caption {The projected phase space diagram for the composite cluster constructed using those objects 
having richness in the range $ 20 \leq \lambda \leq 23.5$. Green lines represent the radially-dependent 
$2.7 \sigma_{\text{LOS}}$ cut used to reject interlopers (indicated by black dots). }
\label{fig:phase_space1}
\end{figure}

\subsection{Interloper rejection}
\label{sec:interloperrejection}

The observables on which the analysis is based are the galaxy projected clustercentric distance R and the rest frame line of sight (LOS) velocity $\mathrm{v}_\text{rf}$. We extract $\mathrm{v}_\text{rf}$ from the galaxy redshift $z_\text{gal}$ and equivalent velocity $\mathrm{v}(z_\text{gal})$ as $\mathrm{v}_{\text{rf}} \equiv (\mathrm{v}(z_\text{gal}) - \mathrm{v}(z_\text{c}))/(1+z_{\text{c}})$, with $z_{\text{c}}$ being the cluster redshift.

Even though the SPIDERS automated procedure assesses membership for each galaxy, there could still be interloper galaxies, i.e. galaxies that are projected inside the cluster virial region, but do not actually lie inside it. To reduce this contamination, we apply the ``Clean" method \citep{2013MAMPOSSt}, which uses the projected phase space location of each galaxy and its comparison to the expected maximal line of sight velocity at each projected radius estimated for the cluster.  Because we do not have enough spectroscopic redshifts to do this accurately for each individual cluster, we divide our sample in bins of richness and perform the interloper rejection in each of them separately.  Specifically, we divide the sample into 15 equally spaced $\lambda$ bins and build a composite cluster in each bin. We apply no scaling in velocity, and stack in physical radius [Mpc] to build the composite clusters.

The ``Clean" method is implemented through several steps. First, the cluster mass is estimated from the LOS velocity dispersion $\sigma_{\text{LOS}}$ of each composite cluster, using a scaling relation calibrated using numerical simulations \citep[e.g.,][]{2013Saro}, and assuming an NFW mass profile with concentration sampled from the theoretical mass--concentration relation of \citet{Maccio2008}.  Thereafter, assuming the \citet[][M{\L}]{2005Mamon} velocity anisotropy profile model, and given the $M(r)$ of the cluster, a Gaussian LOS velocity dispersion profile with $\sigma_{\text{LOS}}(R)$ is calculated and used to iteratively reject galaxies with  $| \mathrm{v}_\text{rf} |  > 2.7 \sigma_{\text{LOS}}$ at any clustercentric distance \citep[see][]{Mamon2010, 2013MAMPOSSt}. In Fig.~\ref{fig:phase_space1} we show the location of galaxies in projected phase space with the identification of cluster member galaxies for the composite cluster constructed using those objects having richness in the range  $ 20 \leq \lambda \leq 23.5$.

\begin{figure}
\centering 
\includegraphics[scale=0.45]{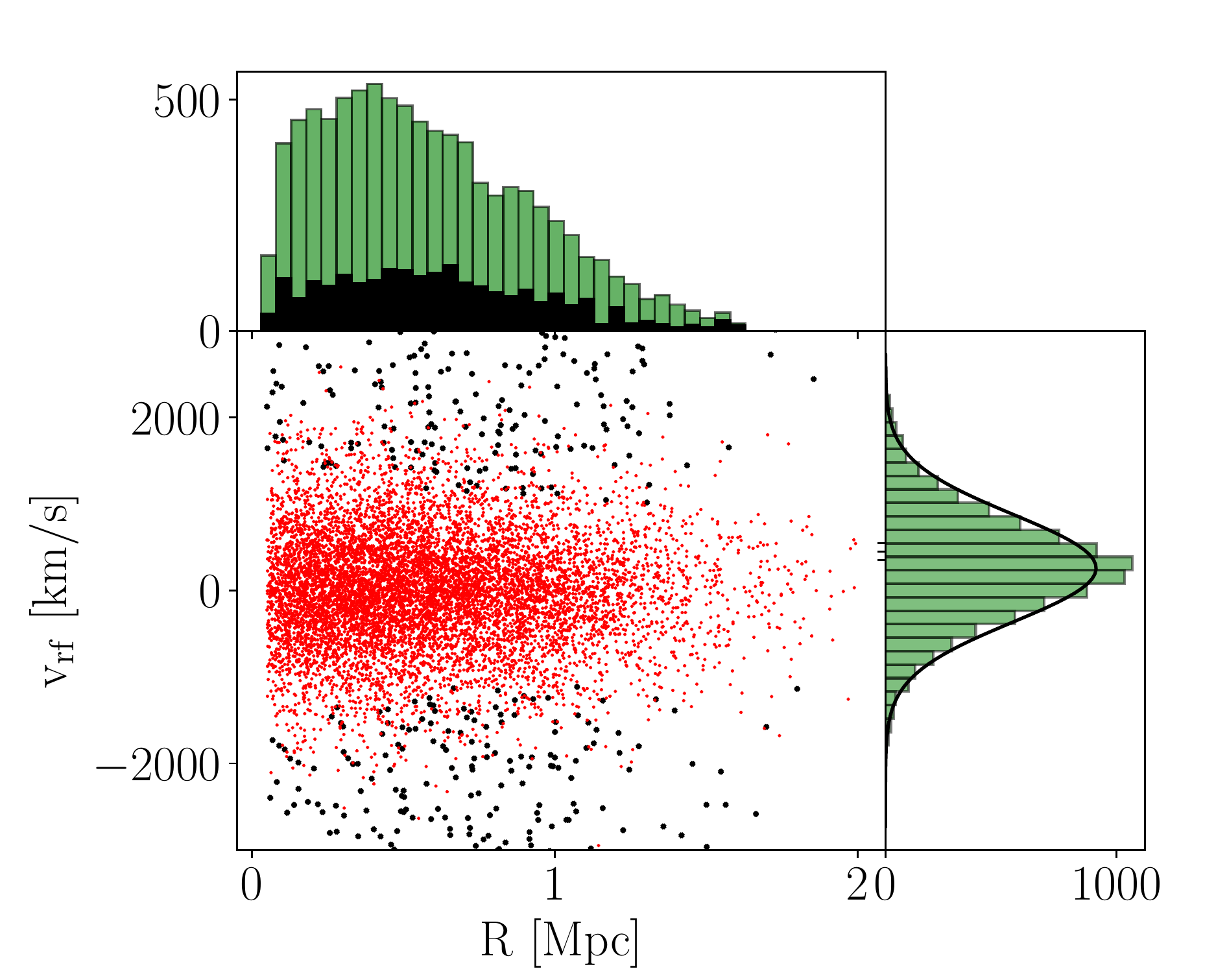}
\caption {Projected phase space distribution for the final sample of 428 clusters. Red dots indicate the 7807 cluster members, while black dots mark the $\sim$2000 rejected interloper galaxies. In the upper panel we show in green the radial distribution of the member galaxies with measured redshifts, and in black the radial distribution of the interlopers. The panel on the right shows the distribution of rest-frame velocities, with an overplotted Gaussian of the same dispersion for comparison.}
\label{fig:phase_space_tot}
\end{figure}

The distribution of the final sample of galaxies in projected phase space is presented in Fig.~\ref{fig:phase_space_tot}. In this plot we show the galaxies identified as cluster members (red dots), the rejected interlopers (black dots), and the radial and velocity distributions of the member galaxies with measured redshifts (green histograms). 

We also note that, even after carrying out interloper rejection, there is still a degree of contamination by interlopers. In fact, galaxies that lie outside the virial radius will tend to have smaller peculiar velocities than those galaxies lying within the virial region.  Indeed, close to the cluster turn-around radius the galaxies will have negligible peculiar velocity and cannot be removed from the sample through an interloper rejection algorithm of the type we adopt here. In fact there is no obvious method for separating these galaxies from the sample within the cluster virial region that we wish to model.  An analysis of cosmological $N$-body simulations carried out by \citet{2013Saro} shows that, when passive galaxies are selected, this contamination is characteristically $\sim$20\% for massive clusters ($M_\mathrm{200c}\geq10^{14}M_\odot$).  For less massive clusters the contamination is expected to be higher.  Another work carried out by \citet{Mamon2010}, based on hydrodynamical cosmological simulations, showed that the distribution of interlopers in projected phase space is nearly universal, presenting only small trends with cluster mass. They state that, even after applying the iterative $2.7 \sigma_{\text{LOS}}$ velocity cut, the fraction of interlopers is still  23 $\pm$ 1\% of all DM particles with projected radii within the virial radius, and over 60\% between 0.8 and 1 virial radius. Further exploration of the effects of this contamination on the dynamical analysis is required, and we are pursuing that in a separate study (Capasso et al., in prep.).

\begin{figure}
\centering 
\includegraphics[scale=0.40]{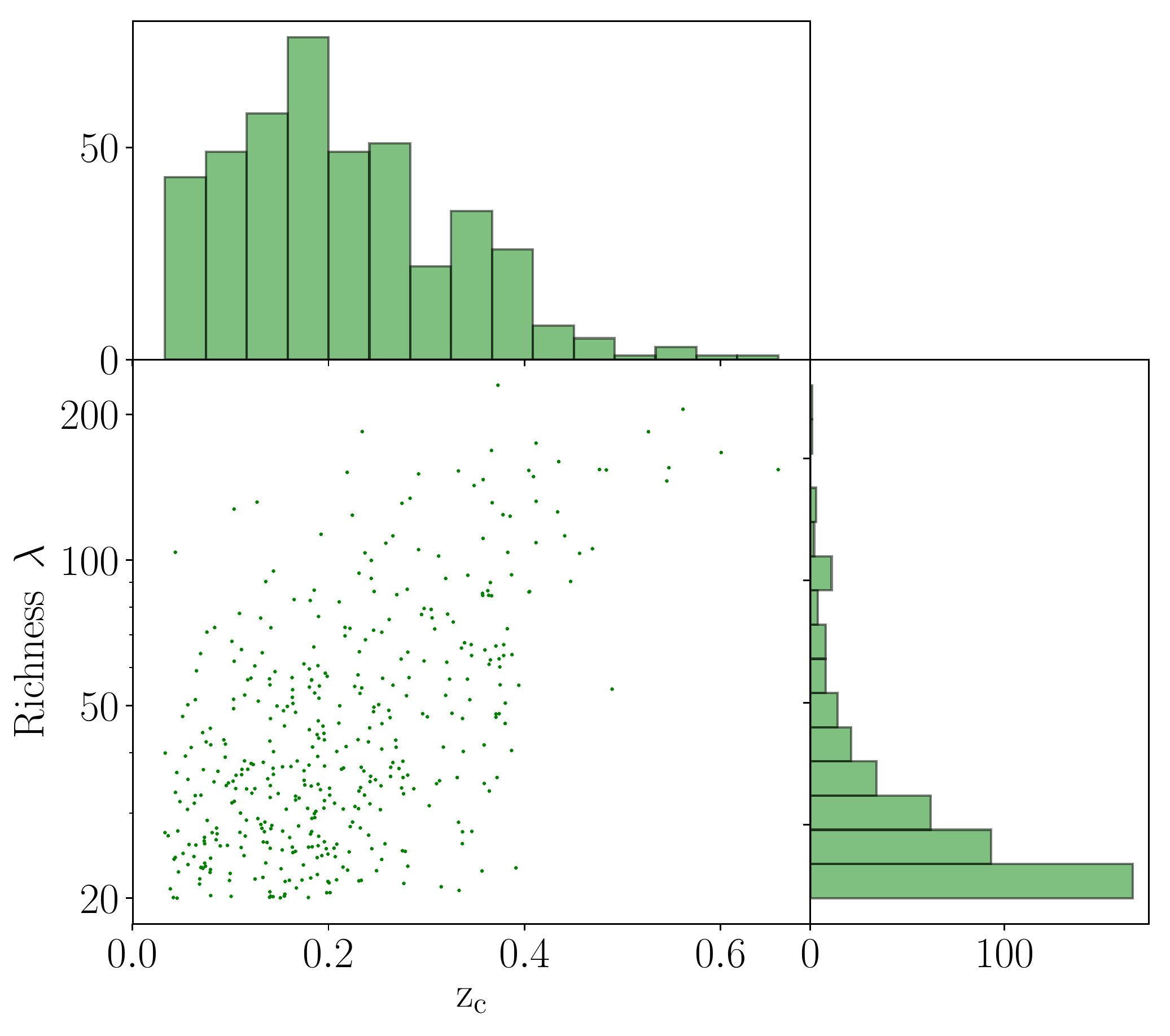}
\caption {Distribution of richness $\lambda$ and cluster redshift $z_{\rm{c}}$ of the final cluster sample.}
\label{fig:histo_rich_z}
\end{figure}

After the application of the interloper rejection, we are left with a total of 703 clusters and 9,121 red galaxies. For the analysis presented here we apply another cut on the cluster sample, keeping all CODEX systems that currently have at least 10 spectroscopic members, $N_{\text{mem}}\ge 10$.   After this cut, our sample consists of 428 clusters and 7807 red galaxies, with a median redshift, richness, and luminosity of z = 0.18, $\lambda$=41, and $\text{L}_{\text{X}}=  9.2 \times 10^{43}$erg\,s$^{-1}$, respectively. Fig.~\ref{fig:histo_rich_z} shows the distributions of cluster redshift and richness of the final sample.

\subsection{Galaxy number density profile}
\label{sec:number_density}

The Jeans analysis requires knowledge of the 3D number density profile $\nu (r)$ of the tracer populations  whose dynamical properties are being used to study the mass and orbital properties of the system.  In our  case, these are the red sequence galaxies selected by the redMaPPer algorithm for observations within SPIDERS.
As only the logarithmic derivative of $\nu (r)$ enters the Jeans equation (see equation~\ref{eq:jeans}), the absolute normalization of the galaxy number density profile has no impact on our  analysis. However, a radially dependent incompleteness in the velocity sample would impact our analysis.  In general, the spectroscopic followup within SPIDERS will lead to a radially dependent incompleteness. This means we cannot simply adopt the spectroscopic sample to measure the number density profile of the tracer  population. We therefore rely on a study of the galaxy populations in  74 Sunyaev-Zel'dovich effect (SZE) selected  clusters from the SPT-SZ survey, which have been imaged as part of the Dark Energy Survey Science Verification phase \citep{2017Hennig}. That study found no mass or redshift trends in the radial distribution of red sequence galaxies for $z>0.25$ and $M_\mathrm{200c}>4\times10^{14}M_\odot$, finding the number density profile of the red sequence population to be well fit by a Navarro, Frenk and White (NFW) model \citep[][]{1996NFW} out to radii of 4$r_\mathrm{200c}$, with a  concentration for cluster galaxies of $c_{\text{gal}} = 5.37^{+0.27}_{-0.24}$. Therefore, we adopt the number density profile described by an NFW profile with the above-mentioned value of $c_{\text{gal}}$ and a scale radius $r_{\text{s, gal}} = R_\mathrm{200c}/c_{\text{gal}}$, making the assumption that the dynamical properties of our spectroscopic sample are consistent with the dynamical properties of the red sequence galaxy population used to measure the radial profiles.

We note that the \citet{2017Hennig} study indicates significant cluster to cluster scatter in the NFW concentration.  We do not expect this scatter to be a source of significant bias in our analysis, because in an earlier analysis \citet{2019Capasso} showed that the mean masses extracted from composite clusters and from the fitting of an ensemble of individual clusters are in good agreement.  We will nevertheless further examine the impact of mismatch between the model and actual radial distribution of the tracer population in an upcoming study where we seek to improve the understanding of biases and scatter in dynamical mass estimators using mock observations of structure formation simulations (Capasso et al., in prep.).


\section{Theoretical Framework}
\label{sec:theory}

The method we adopt for the dynamical analysis of our clusters is based on the spherically-symmetric Jeans analysis \citep{1987Binney}. Using the Jeans equation, it is possible to define the mass distribution $M(r)$ of a cluster as
\be                                                             
\label{eq:jeans}                                                              
\frac{GM(<r)}{r} = - {\sigma_{r}^{2}} \left( \frac{d \ln \nu}{ d \ln r } + \frac{d \ln {\sigma_{r}^{2}}} { d \ln r} + 2 
\beta \right) ,
\ee
where $\nu(r)$ is the number density profile of the tracer galaxy population, $\sigma_{r}(r)$ is the radially dependent component of the velocity dispersion along the spherical coordinate $r$, $M(<r)$ is the enclosed mass within radius $r$, $G$ is Newton's constant, $\beta(r) \equiv 1 - (\sigma_{\theta}^{2} / \sigma_{r}^{2})$ is the radially dependent velocity dispersion anisotropy, and $\sigma_{\theta}$ is the tangential component of the velocity dispersion. 
The observables we employ to constrain these quantities are projected quantities, including the surface density profile of the galaxy distribution, the rest frame LOS velocities and the radial separation of each galaxy from the cluster center.

Given the limited knowledge of the line of sight velocity distribution within realistic cluster dynamical datasets, it is not possible to uniquely derive the mass distribution of a galaxy cluster \citep{1987Merritt}. To address this problem, we use the Modeling Anisotropy and Mass Profiles of Observed Spherical Systems algorithm \citep[hereafter MAMPOSSt; for full details please refer to][]{2013MAMPOSSt}. This code performs a maximum likelihood analysis of the projected phase space distribution of the observed sample using the theoretical distribution predicted for a given model using the Jeans equation.  The observations are used to constrain the model parameters adopted to describe the cluster mass distribution and galaxy orbital anisotropy. The MAMPOSSt method thus requires adopting parametrized models for the number density, mass, and velocity anisotropy profiles $\nu(r)$, $M(r)$, $\beta(r)$. As addressed in Section~\ref{sec:number_density}, because our spectroscopic dataset might suffer from radially dependent incompleteness, we adopt the measured number density profile derived from the study of red sequence galaxies in SZE selected clusters \citep{2017Hennig}.  We discuss our choice of the mass and velocity anisotropy profiles in the next section.

\subsection{Mass and anisotropy profiles}
\label{sec:profiles}

Taking guidance from both numerical studies of structure formation and observational results, we adopt the mass model 
introduced by \citet[][NFW]{1996NFW}
\be
\rho(r)=\rho_0 \left({r\over r_\mathrm{s}}\right)^{-1}\left(1+{r\over  r_\mathrm{s}}\right)^{-2},
\ee
where $\rho_0$ is the central density, and $r_\mathrm{s}$ is the scale radius where the logarithmic derivative of the density profile reaches -2. Integrating this density profile up to $r_\mathrm{200c}$, we obtain the mass enclosed inside the virial radius
\be
M_{\text{200c}}=4\uppi\rho_{0}r_\mathrm{s}^{3}\left[\ln\left(\dfrac{r_\mathrm{s} + r_\mathrm{200c}}{r_\mathrm{s}}
\right) -\dfrac {r_\mathrm{200c}}{r_\mathrm{s} + r_\mathrm{200c}}\right]  .
\ee
Cosmological simulations produce dark matter halos with mass profiles well described by this profile. Even though some results have preferred different models \citep{2006Merritt, 2010Navarro, 2014Dutton, 2015vdb, 2017Sereno}, this result is in good agreement with a variety of observational analyses using both dynamics and weak lensing \citep{1997Carlberg, vanderMarel2000, 2003Biviano, Katgert2004, 2011Umetsu}.

For the velocity anisotropy profile, we consider five models that have been used in previous MAMPOSSt analyses and that are described also in \citet{2019Capasso}.  These are (1) the constant anisotropy model (C), (2) the Tiret anisotropy profile \citep[][T]{2007Tiret}, (3) the \citet{2005Mamon} profile (M{\L}), (4) the Osipkov-Merritt  anisotropy profile \citep[][OM]{1979Osipkov,1985Merritt}, and (5) a model with anisotropy of opposite sign at the center and at large radii (O). 

Therefore, to predict the projected phase space distribution of the observed dynamical dataset for each  cluster, we run MAMPOSSt with 3 free parameters: the virial radius $r_\mathrm{200c}$, the scale radius $r_\mathrm{s}$ of the mass distribution, and a velocity anisotropy parameter $\theta_{\beta}$. This parameter represents the usual $\beta=1-(\sigma_{\theta}^{2}/\sigma_{r}^{2})$ for the first three models (C, T, O), while for the $\text{M{\L}}$ and OM models it defines a characteristic radius $\theta_\beta=r_\beta$.  

\subsection{Bayesian model averaging}
\label{sec:Bayes}

As the literature does not provide us with strong predictions for the radial form of the velocity anisotropy profile $\beta(r)$, we employ all the five models described above when estimating the cluster masses. We combine the results from the different models by merging their constraints exploiting the Bayesian model averaging technique. A weight is assigned to each model, which is proportional to how well the model fits the data. This weight is represented by the so-called Bayes factor \citep[see][and references therein]{Hoeting99bayesianmodel}. 

Considering the 5 anisotropy models $M_{1}$, ..., $M_{5}$, we define the Bayes factor $B_{j}$ of each model $j$ by normalizing the marginalized likelihood of the model $\mathcal{L}(D \,| M_{j})$, also known as evidence, by the likelihood of the most probable model.  Specifically, 
\be
\label{eq:bayesfactor}
B_{j} = {\mathcal{L}(D \,| M_{j}) \over \mathcal{L}(D \,| M_{\text{max}})}  ,
\ee
where $M_{\text{max}}$ indicates the model with the highest marginalized likelihood, $\mathcal{L}(D \,| M_{j}) = \int { \mathcal{L}(D\,| \theta_{j}, M_{j}) P(\theta_{j}\,| M_{j}) \,\, d \theta_{j} }$, $\mathcal{L}(D\,| \theta_{j}, M_{j})$ is the likelihood of the data $D$ given the model parameters $ \theta_{j}$, and $P(\theta_{j}\,| M_{j})$ is the prior. 

The average posterior distribution on the parameter common to all anisotropy models is then simply given by the weighted average of the posterior distributions of each model, with the Bayes factor as weight. To perform this Bayesian model averaging, we employ the multimodal nested sampling algorithm MultiNest \citep{2008Feroz, 2009Feroz, 2013Feroz}, which provides us with the evidence for each model.

\begin{table} \centering
\caption{Priors assumed for our analysis. $\mathcal{U}(i, j)$ refers to a uniform flat prior in the interval $(i, j)$, while $\mathcal{N}(\mu, \sigma^{2})$ indicates a Gaussian distribution with mean $\mu$ and variance $\sigma^{2}$.} 
\begin{tabular}{ccccc}
\hline\\[-7pt]
$\mathrm{A}_\lambda$ & $\mathrm{B}_\lambda$ & $\gamma_\lambda$  & $r_{\beta}$ & $\sigma_{\ln\lambda}^{\text{int}}$ \\ [2pt]
\hline\\[-7pt]
$\mathcal{U}(20, 50)$ & $\mathcal{U}(0.5, 2)$ & $\mathcal{U}(-3, 2)$  &  $\mathcal{U}(0.01, 10)$ & $\mathcal{N}(0.15, 0.09^2)$ \\ [3pt]
\hline
\end{tabular}
\label{tab:priors}
\end{table}

\begin{table*} \centering
\caption{ RedMaPPer Richness-mass-redshift scaling relation parameters from this analysis and the literature.  The results from our analysis include corrections for the Eddington and Malmquist biases.  Parameters are defined in equation~(\ref{eq:lambda-mass}).  For results from this analysis the uncertainties are statistical, and a systematic mass uncertainty of 10\% is applied to the amplitude  $\mathrm{A}_\lambda$.  In the comparison to previous results, the amplitude $\mathrm{A}_\lambda$ column contains the $\lambda$ at $M_\mathrm{200c}=3\times10^{14} M_\odot$ and  $z=0.18$.  Conversions have been made to $M_\mathrm{200c}$ and from $E(z)$ to $(1+z)$ where needed. Note also that each of these studies was performed on a different range of mass and redshift.} 
\begin{tabular}{lccc}
\hline\\[-7pt]
Dynamical analyses using SPIDERS data & $\mathrm{A}_\lambda$ & $\mathrm{B}_\lambda$ & $\gamma_\lambda$  \\ [2pt]
\hline\\[-7pt]
Baseline analysis: $\lambda\ge20$, $N_\mathrm{mem}\ge10$ & $38.6^{+3.1}_{-4.1}\pm3.9$ & $0.99^{+0.06}_{-0.07}\pm0.04$  &  $-1.13^{+0.32}_{-0.34}\pm0.49$ \\ [3pt]
As above, but with correlated scatter correction & $39.8^{+3.0}_{-3.8}\pm4.0$ & $0.98^{+0.07}_{-0.07}\pm0.04$  &  $-1.08^{+0.31}_{-0.34}\pm0.49$ \\ [3pt]
\hline\\[-7pt]
Previously published results &$\lambda(3\times10^{14}M_\odot,0.18)$ & $M_\mathrm{200c}^\mathrm{B_\lambda}$ & $(1+z)^{\gamma_\lambda}$ \\[4pt]
\hline\\[-7pt]
WL masses using DES Y1 \citep{2018McClintock} & $43.8\pm1.3$ & $0.73\pm0.03$  &$-0.10\pm0.10$ \\ [3pt]
CMB WL masses \citep{2018Baxter}   & $49.8\pm10.8$ & $0.81\pm0.21$ & -- \\ [3pt]
WL masses using SDSS \citep{2017Simet}  & $63.1\pm2.2$ & $0.74\pm0.06$ & --  \\ [3pt]
Cluster clustering using SDSS \citep{2016Baxter} & $37.5\pm4.4$ & $0.84\pm0.12$ &  $0.70\pm0.90$  \\ [3pt]
Pairwise velocity dispersion with SDSS \citep{2016Farahi} & $47.7\pm1.0$ & $0.75\pm0.04$ & --  \\ [3pt]
SPT masses with RM from DES SV \citep{2015Saro}  & $36.1\pm9.1$ & $1.16\pm0.20$ &  $0.60\pm0.63$ \\ [3pt]
\hline
\end{tabular}
\label{tab:results}
\end{table*}


\section{Results}
\label{sec:results}

In this section we present the results of our dynamical analysis.  In the first subsection we describe how we calibrate the $\lambda$-mass relation and present our results. In the following subsection we explore the impact of correlated scatter in the X-ray luminosity and richness for the CODEX sample. Afterwards, we compare our findings to previous works, and we test how strongly the number of member galaxies per cluster affects our results.


\begin{figure}
\centering { 
   \includegraphics[scale=0.42]{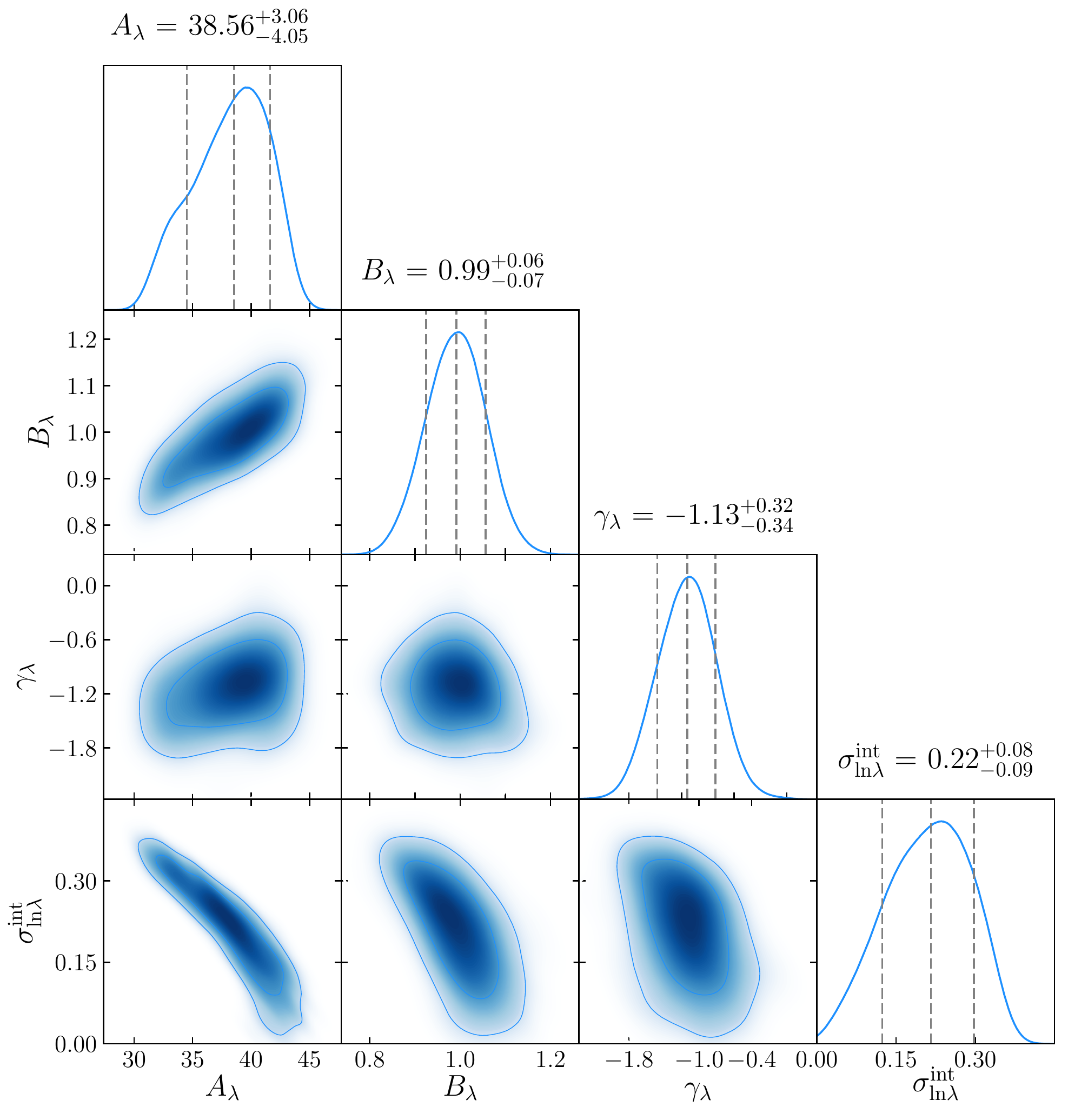} 
      }
      \vskip-0.2in
\caption{Parameters of the $\lambda$-$M_\mathrm{200c}$-$z$ relation. Contours show the 1$\sigma$, 2$\sigma$, and 
3$\sigma$ confidence regions.}
\label{fig:rich_corner}
\end{figure}

\subsection{$\lambda$-$M_\mathrm{200c}$-$z$ relation}
\label{sec:scalingrelation}

We adopt a power-law relation between cluster richness $\lambda$, mass and redshift of the form
\be
\label{eq:lambda-mass}
\lambda = \text{A}_{\lambda} \left(  \dfrac{M_{\text{200c}}} {M_{ \text{piv}}} \right)^{\text{B}
_{\lambda}} \left( \dfrac{1+z}{1+z_{ \text{piv}}} \right)^{\gamma_{\lambda}}, 
\ee
where $\text{A}_{\lambda}$, $\text{B}_{\lambda}$, and $\gamma_{\lambda}$ are the amplitude, the mass slope and the redshift evolution slope.  Similar forms have been used to study the galaxy halo occupation number and richness previously \citep{2004Lin,2006Lin,2015Saro,2017Hennig,2017Saro}.  We adopt the redshift scaling $(1+z)^\gamma$ instead of $E(z)^\gamma$ because, as discussed in a recent study of X-ray scaling relations \citep{2018Bulbul}, we wish to avoid ascribing cosmological sensitivity to redshift trends unless there is a physically justifiable reason to do so.  Sensitivity of an observable to the evolving critical density of the Universe would justify an $E(z)$ scaling.  An example would be an observable like the X-ray luminosity or Sunyaev-Zel'dovich effect signature that depends on the intracluster medium density, but in the case of the galaxy richness or halo occupation number, the density plays no role and no such sensitivity is expected.  We set the pivot redshift to be $z_{ \text{piv}} = 0.18$, which is the median redshift of our cluster sample. We have adjusted the value of the mass pivot $M_{ \text{piv}}=3 \times 10^{14} M_{\odot} $ iteratively to minimize the false degeneracy between $\text{A}_{\lambda}$ and $\text{B}_{\lambda}$.

We marginalize over the intrinsic scatter in $\lambda$ at fixed mass, which is set to be log-normal with a prior on the scatter from \citet{2017Saro}, $\sigma_{\ln\lambda}^{\text{int}} = 0.15 \pm 0.09$ (precise priors listed in Table~\ref{tab:priors}).
We assume the full scatter in $\lambda$ at fixed mass is log-normal with variance given by:
\be
\label{eq:scatter_rich}
\sigma_{\ln\lambda}^{2} = \dfrac{\eta}{\lambda} +  {\sigma_{\ln\lambda}^{\text{int}}}^{2} ,   
\ee
where $\eta$ is the scale factor described in Section~\ref{sec:CODEX} that is a correction factor that accounts for the limited depth of the SDSS photometry in accounting for the richness calculated over a fixed portion of the cluster galaxy luminosity function.

For each cluster $i$ in our sample, we calculate an initial mass $M_{\text{200c,obs}}$ using the scaling relation described in equation~(\ref{eq:lambda-mass}) and the current values of the parameter vector $\textbf{\textit{p}}$, which contains the 4 scaling relation parameters $A_{\lambda}, B_{\lambda}, \gamma_{\lambda}$, and $\sigma_{\ln\lambda}^{\text{int}}$ together with the anisotropy model parameter $r_{\beta}$.  In each iteration we use the current value of the scatter $\sigma_{\ln\lambda}$ to estimate a correction for the Eddington bias caused by the interplay of the $\lambda$ scatter and the mass function using the method described in \citet{2011Mortonson}. Assuming a log-normal mass observable relation with variance $\sigma^2_{\ln{M}} = (1 / B_{\lambda} \cdot \sigma_{\ln\lambda})^{2} $ that is small compared with the scale over which the local slope $\Gamma$ of the mass function changes, the posterior mass distribution is a log-normal of the same variance $\sigma^2_{\ln{M}}$ with a shifted mean $\ln <M_{\text{200c,true}}> = \ln <M_{\text{200c,obs}}> + \Gamma\sigma_{\ln M}^{2} $. 

With this mass, we then use MAMPOSSt to construct the probability distribution in projected phase space for each cluster, combining the likelihoods calculated for each member galaxy in that cluster
\be
\label{eq:lik_rich}
\mathcal{L}_{i}=\prod_{j \in gal} \, \mathcal{L} (R^{j}, v^{j}_{rf}, \lambda_{i}, z_{i} \mid  \textbf{\textit{p}}),
\ee
where $R^{j}$ and $v^{j}_{rf}$ are the clustercentric radii and rest-frame velocities of the member galaxy $j$ in the cluster $i$.  The maximum likelihood solutions are obtained using the \textsc{newuoa} software \citep{newuoa}. Flat priors are assumed for the scaling relation parameters $\text{A}_{\lambda}$, $\text{B}_{\lambda}$, and $\gamma_{\lambda}$, and for the anisotropy parameter $r_{\beta}$ (see Table~\ref{tab:priors}).

We combine the likelihoods for all these clusters, to then obtain the likelihood for the total sample for each set of scaling relation parameters~$\textbf{\textit{p}}$, i.e. $\mathcal{L}  = \prod_{i \in clus} \mathcal{L}_{i} $. This procedure must be done separately for each anisotropy profile model (see Section~\ref{sec:profiles}). Finally, we use Bayesian model averaging to combine the posterior parameter distributions obtained from the different anisotropy models, effectively marginalizing over the uncertainties in the orbital anisotropy.

Because we impose a cut on our observable, $\lambda \ge 20$, a correction for the Malmquist bias is also needed \citep{2000Sandage}. We estimate the effect of this correction by creating a large mock catalogue ($\sim4400$ clusters and $\sim165,500$ member galaxies) by computing the number of expected clusters as a function of halo mass and redshift using the halo mass function \citep{2008Tinker}. We then draw a Poisson realization of the number of expected clusters, obtaining a mass selected cluster sample with $M_{\text{200c}} \ge 7\times 10^{13} $ and $ 0.05 \le z \le 0.66 $. 
Using the scaling relation parameters recovered from our analysis before correcting for this bias, we calculate $\lambda$ for each cluster of mass $M_{\text{200c}}$.  Scatter is added to this relation such that the assigned $\lambda$ values are sampled from a Gaussian distribution having scatter given by equation~(\ref{eq:scatter_rich}). The mock sample we produce has richnesses $\lambda > 6.5$.  
For each cluster in our mock sample, we create a sample of member galaxies. We run MAMPOSSt on a grid of velocities and clustercentric distances, fixing the galaxy number density profile to that described in Section~\ref{sec:number_density}, and generating a random number of galaxies per cluster drawn from the distribution of member galaxies in our observed sample. Finally, we use the MAMPOSSt likelihood to recover the probability density of observing an object at a certain location in phase space \citep[see equation~11,][]{2013MAMPOSSt}. 

We fit this mock dataset and recover best fit parameter values that are consistent with the input values.  Then we impose a $\lambda>20$ richness cut on the sample and refit, noting that the best fit mass and redshift trends are affected.  Using this approach, we estimate corrections for the Malmquist bias that correspond to $\delta B_{\lambda}$=+0.05 and $\delta \gamma_{\lambda}$=-0.06. These corrections are included in all the results we present. 

Table~\ref{tab:results} summarizes the posterior of our model parameters from our so-called ``baseline analysis", i.e. before accounting for the impact of correlated scatter (see Section~\ref{sec:correlatedscatter}), while Fig.~\ref{fig:rich_corner} shows the corresponding joint parameter constraints. Our results imply that galaxy clusters with redshift $z = 0.18$ and mass $M_{\text{200c}} = 3 \times 10^{14} M_{\odot}$ have a mean richness of $\text{A}_{\lambda} = 38.56^{+3.06}_{-4.05}$. The mass scaling is consistent with linear, $\text{B}_{\lambda} = 0.99^{+0.06}_{-0.07}$. The redshift dependence in the CODEX sample is $\gamma_{\lambda} = -1.13^{+0.32}_{-0.34}$, indicating that the red sequence richness $\lambda$ at fixed mass falls as one moves to higher redshift.

 
 \subsection{Additional Systematic Effects}
 \label{sec:systematics}
 
The results presented in the last section include corrections for the Eddington bias and the Malmquist bias, but the uncertainties on the parameters reflect only statistical errors.  In this section we consider systematic effects and the impact they have on the best fit parameters and the parameter uncertainties.  

We estimate that there is an additional 10\% systematic uncertainty associated with the dynamical mass measurements themselves. This estimate comes from an analysis of the MAMPOSSt code run on numerical simulations in the analysis of \citet[][]{2013MAMPOSSt}. In their work, the authors show that, using particles lying within a sphere of $r_{100}$ around the halo center, the estimate of the cluster virial radius $r_\mathrm{200c}$ is biased at $ \leq 3.3\%$ \citep[see Table 2,][]{2013MAMPOSSt}. Therefore, we adopt a Gaussian systematic uncertainty on the virial mass $M_\mathrm{200c}$ of $\sigma = 10\%$.  The \citet{2013MAMPOSSt} analysis does not explore mass or redshift trends in these biases, and therefore we apply the entire uncertainty to the normalization parameter $\mathrm{A}_\lambda$.  In a future analysis, we plan to explore the mass and redshift dependence of the systematic uncertainties in dynamical mass estimates from a Jeans analysis (Capasso et al., in prep.). 
 
In the subsections below we first consider the impact of selecting different subsamples using the number of member galaxies with spectroscopic redshifts $N_\text{mem}$, and then we explore the impact of possible correlated optical and X-ray scatter.


\begin{table} 
\centering
\caption{Impact of the number of spectroscopic members on the redMaPPer Richness-mass-redshift scaling relation parameters. Parameters are defined in equation~(\ref{eq:lambda-mass}).  The uncertainties on the results are statistical, corresponding to 68 per cent confidence intervals, and a systematic mass uncertainty of 10\% is applied to the amplitude  $\mathrm{A}_\lambda$.} 
\begin{tabular}{cccc}
\hline\\[-7pt]
Number of cluster  & $\mathrm{A}_\lambda$ & $\mathrm{B}_\lambda$ & $\gamma_\lambda$  \\ [2pt]
member galaxies &  & &   \\ [2pt]
\hline\\[-7pt]
$N_\mathrm{mem}\ge1$ & $39.2^{+2.9}_{-3.5}$ & $0.91^{+0.05}_{-0.06}$  &  $-0.15^{+0.23}_{-0.24}$ \\ [3pt]
$N_\mathrm{mem}\ge3$ & $39.3^{+3.1}_{-3.6}$ & $0.92^{+0.05}_{-0.06}$  &  $-0.26^{+0.23}_{-0.24}$ \\ [3pt]
$N_\mathrm{mem}\ge5$ & $39.2^{+3.0}_{-3.7}$ & $0.95^{+0.06}_{-0.06}$  &  $-0.65^{+0.26}_{-0.27}$ \\ [3pt]
$N_\mathrm{mem}\ge10$ & $38.6^{+3.1}_{-4.1}$  & $0.99^{+0.06}_{-0.07}$  &  $-1.13^{+0.32}_{-0.34}$  \\ [3pt]
$N_\mathrm{mem}\ge20$ & $41.6^{+2.5}_{-3.2}$ & $0.98^{+0.09}_{-0.08}$  &  $-1.00^{+0.49}_{-0.56}$ \\
\hline
\end{tabular}
\label{tab:Nmem}
\end{table}

\begin{figure}
\centering 
\includegraphics[scale=0.55]{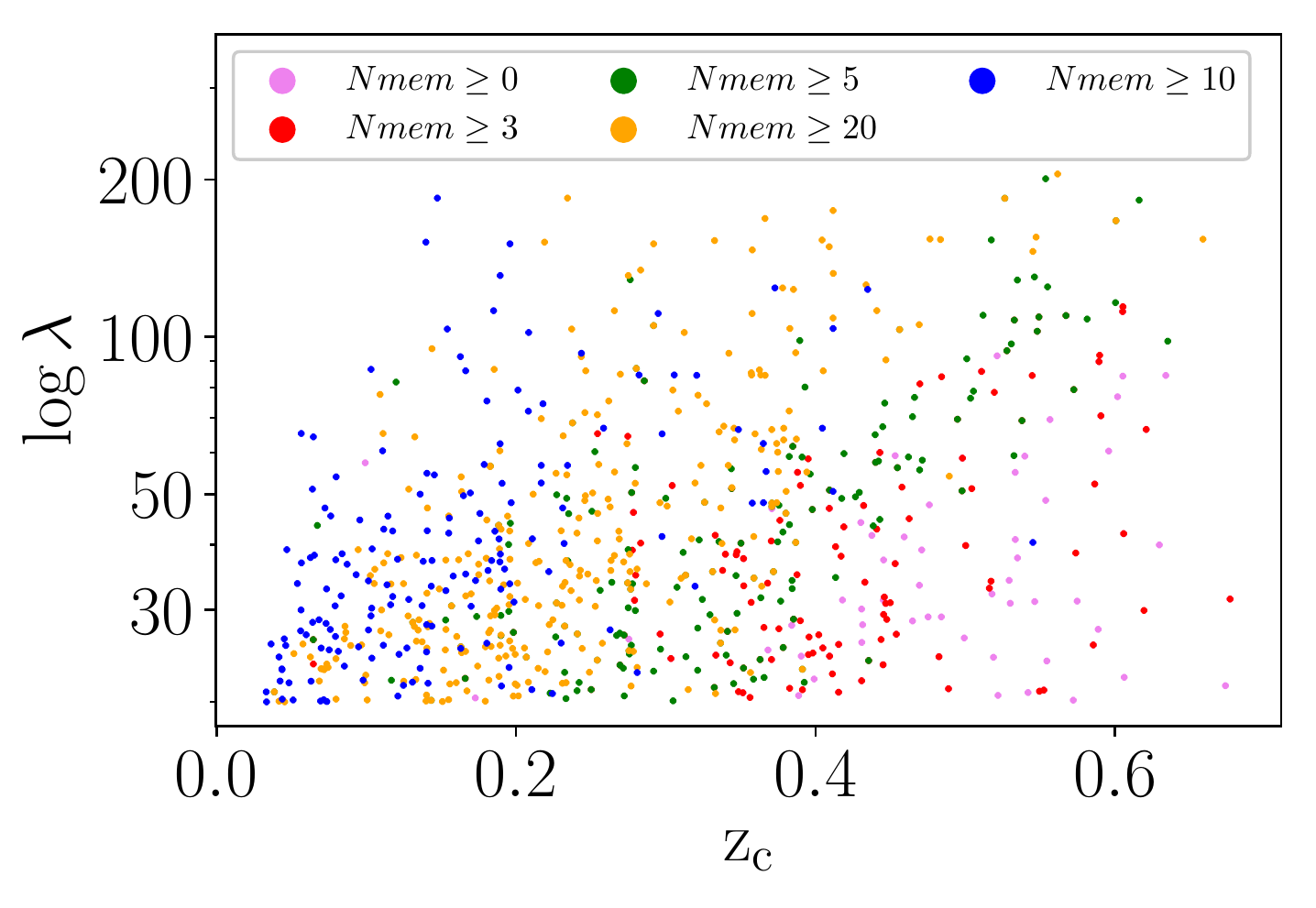}
\caption {Richness and redshift distribution of clusters having a different number of spectroscopic members.}
\label{fig:Nmem}
\end{figure}

\subsubsection{Impact of number of cluster member galaxies}
\label{sec:memcut}

As described in Section~\ref{sec:finalmemberselection}, we apply a cut to our sample prior to the dynamical analysis, keeping only those systems having at least 10 spectroscopic members: $N_{\text{mem}}\ge10$. This decision is driven by our concern that good constraints on the cluster masses and scaling relation parameters could not be obtained from clusters having very small numbers of spectroscopic members.  However, this selection is somewhat arbitrary, and so we explore here the impact of varying this cut.

Table~\ref{tab:Nmem} shows the results obtained imposing different cuts on the number of spectroscopic members, where $N_{\text{mem}}$ varies from 1 to 20.  Note that the BCG has been excluded, so the clusters with a single galaxy actually have two measured redshifts.  Interestingly, the normalization $\mathrm{A}_\lambda$ and the mass trend parameter $\mathrm{B}_\lambda$ are not significantly affected when analyzing clusters having a different number of spectroscopic members.  

On the other hand, the value of the redshift trend parameter $\gamma_\lambda$ varies considerably, even reaching values consistent with zero evolution when including clusters having $N_{\text{mem}}\ge1$ and $N_{\text{mem}}\ge3$. The value of $\gamma_\lambda$ becomes stable when including only clusters with at least 10 spectroscopic members, justifying our decision of including only those clusters into our main analysis.  However the strong dependence of $\gamma_\lambda$ on the member cut is an indication of remaining systematic uncertainties on this parameter.

The reason of the different behavior of $\gamma_\lambda$ with respect to that of $\mathrm{A}_\lambda$ and  $\mathrm{B}_\lambda$ is clarified to some degree in Fig.~\ref{fig:Nmem}, where we show the distribution in richness and redshift of galaxy clusters having a different number of spectroscopic members.  The distribution of clusters having $N_{\text{mem}}<10$ extends to higher redshifts, allowing for improved constraints on the redshift trend and also introducing a qualitatively different population of clusters into the analysis.

As the spectroscopic sample at these higher redshifts is increased, we will begin to see whether the trend in $\gamma_\lambda$ with the $N_{\text{mem}}$ cut is revealing a systematic in dynamical masses in the limit of very low spectroscopic sampling of each halo or whether the weaker trends shown with the less dramatic cuts that then include more high redshift systems is really a reflection of the true redshift trend in the $\lambda$-mass relation.  But at this point we use the trend in $\gamma_\lambda$  that is apparent in Table~\ref{tab:Nmem} to estimate a systematic uncertainty on that parameter.  Specifically, we adopt half the full range of variation in the value as the systematic uncertainty on the parameter $\sigma_{\text{sys},\gamma_\lambda}={\Delta |\gamma_\lambda |\over 2}=0.49$.  Similarly for the mass trend parameter we estimate $\sigma_{\text{sys},B_\lambda}={\Delta |B_\lambda |\over 2}=0.035$.  For the amplitude parameter $A_\lambda$ the shift is small compared to the 10\% systematic uncertainty described at the beginning of this section. These systematic uncertainties are listed in Table~\ref{tab:results}.
 
\subsubsection{Impact of correlated $\lambda$ and $L_\mathrm{X}$ scatter}
\label{sec:correlatedscatter}

Before comparing our results to those from the literature, we examine the impact of correlated scatter in the richness and X-ray luminosity on the parameters of the richness mass relation.  To do this we employ the selection function of the CODEX survey calculated as described below by the CODEX team.

As described above in Section~\ref{sec:CODEX}, the CODEX cluster catalog is based on the identification of faint X-ray sources with the help of redMaPPER follow-up on the SDSS photometry to identify optical counterparts.  The final catalog is therefore subject to both X-ray and optical selection in a manner that has been modeled based upon several observational results.
First, the LoCuSS survey \citep[Local Cluster Substructure Survey][]{2010Okabe, 2018Haines} indicates a negative value of the covariance at fixed mass of the scatter in the X-ray luminosity $\text{L}_{\text{X}}$ and the optical richness $
\lambda$. For the selection function modeling adopted here,  the covariance coefficient is fixed to be $\rho_{\mathrm{L}_\mathrm{X}-\lambda}=-0.2$ (Farahi et al. submitted). The net effect of this correlated scatter is that the CODEX survey is more sensitive in detecting clusters of given mass if they have lower richness, because that lower richness is correlated to a higher X-ray luminosity.   The modeling of the survey selection function takes into account the covariance of the scatter in $\text{L}_\text{X}$-mass relation with the shape of the cluster, which affects the sensitivity to a particular cluster. In modeling the selection function, the scaling relations are fixed to those of the XXL survey \citep[e.g.][]{2016Pacaud}, which is well suited for our study here, because it includes both cluster and group mass scales.  

Using the selection function described above, the CODEX team then estimated the effective solid angle of the CODEX survey as a function of the scatter in $\lambda$ as a function of redshift and mass.  The idea here is that because scatter to lower $\lambda$ is weakly correlated to an increase in the cluster  X-ray luminosity, one is effectively probing a larger solid angle for those clusters with lower than typical  $\lambda$ at each redshift and mass.  It is with this data product that we begin our analysis. 

\begin{figure}
\centering 
\includegraphics[scale=0.55]{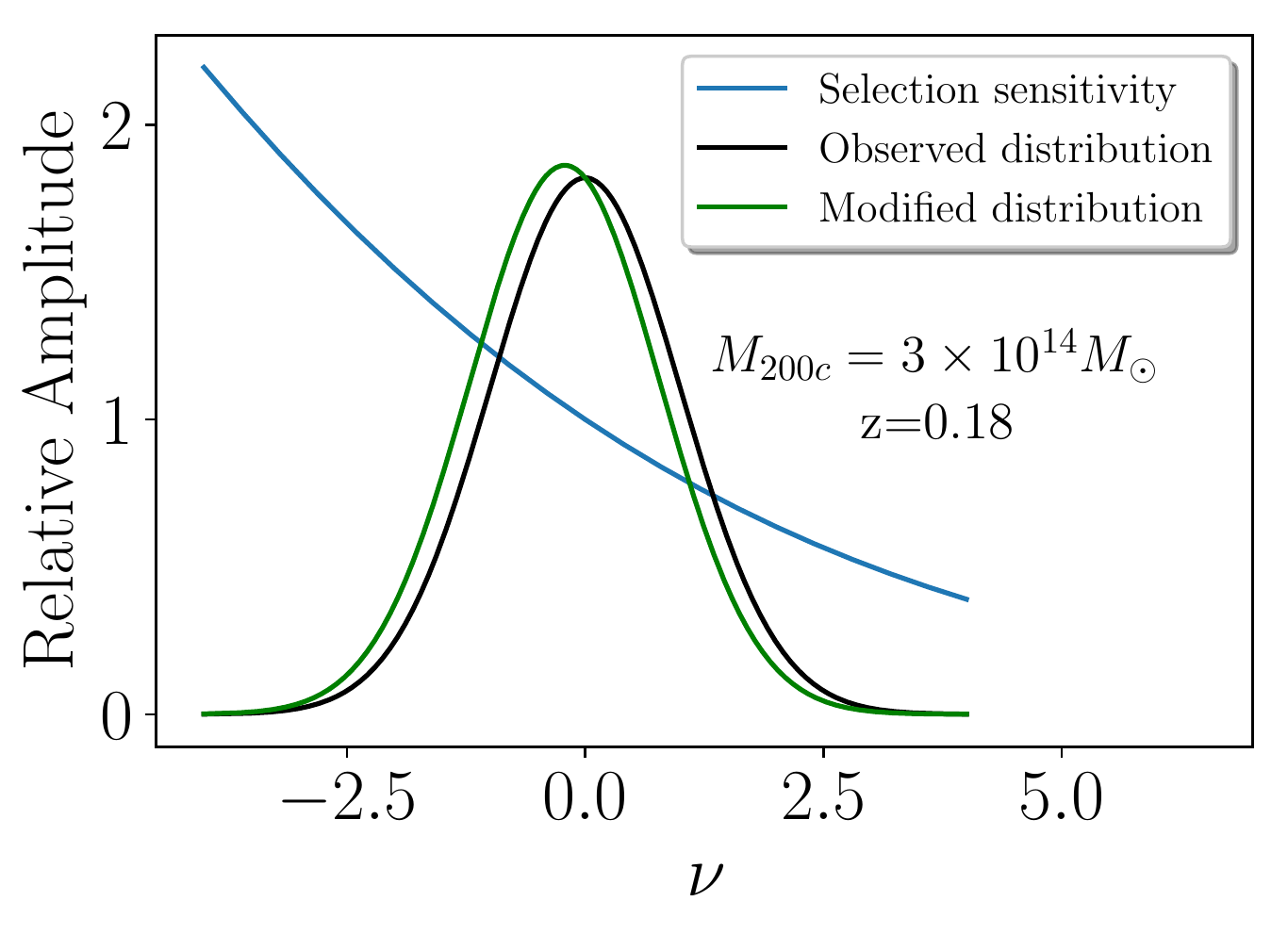}
\caption {Effect of the selection function on the $\lambda$ distribution. In blue we show the the relative sensitivity of the CODEX X-ray selected sample as a function of deviation from the mean observed $\lambda$, normalized to its value at $\nu=0$. The black curve shows the distribution of observed $\lambda$, as a function of deviation $\nu$ from the mean value, while the green distribution shows how the inclusion of the selection sensitivity causes a shift and distortion of the observed $\lambda$ 
distribution. }
\label{fig:matrix_distr}
\end{figure}

To estimate the impact of this correlated scatter on our results, we calculate its effects {\it a posteriori}, using the results of our baseline analysis as listed in Table~\ref{tab:results}.  The variation in sensitivity as a function of $\lambda$ at fixed mass and redshift produces a modification in the shape of the richness distribution at each mass and redshift. In Fig.~\ref{fig:matrix_distr} we show 
an example of how this affects the cluster distribution in $\lambda$ at $M_{\text{200c}} = 3\times10^{14} M_{\odot}$ and $z=0.18$. The blue line represents the relative sensitivity $s(\nu)$ of the CODEX X-ray selected sample as a function of the deviation $\nu$ from the mean, expected $\lambda$ (expressed in equation~\ref{eq:lambda-mass}). This deviation is defined as a function of
\be 
\nu=\frac{\Delta \ln \lambda}{\sigma_{\ln \lambda}^{\text{int}}},
\ee
and the sensitivity function has been normalized to its value at $\nu=0$. The black curve shows the log-normal parent distribution of $\lambda$ at this mass and redshift (equation~\ref{eq:scatter_rich}), as a function of the deviate $\nu$.  
In this space this distribution is simply a Gaussian of unit width.  The green distribution shows the product of the parent $\lambda$ distribution with the selection sensitivity.  Given the $\nu$ dependence of the sensitivity, the new $\lambda$ distribution is well approximated as being a new log-normal distribution with mean shifted away from zero. The shift in the parent $\lambda$ distribution can be written
\be
\left<\nu  | M_\text{200c},z\right> = \int d\nu\, P(\nu)\,s\left (\nu |M_\text{200c},z\right).
\ee
where $P(\nu)$ is the parent $\lambda$ distribution (log-normal) and $s\left (\nu |M_\text{200c},z\right)$ is the sensitivity as a function of $\nu$ given the cluster mass and redshift.  For the given example, the mean shift is $\left<\nu  | M_\text{200c},z\right>=-0.20$.  This shift changes little with mass, but it does evolve with redshift.  This fractional logarithmic shift then implies a shift in $\lambda$ for any given mass and redshift 
\be
\lambda_\mathrm{cor} =\lambda\left(M_\text{200c},z\right) e^{-\left<\nu | M_\text{200c},z\right> \sigma_{\ln \lambda}}
\ee

To estimate the impact on the scaling relation parameters, we calculate $\lambda_\mathrm{cor}$ over the full range of $M_\text{200c}$, $z$ where we have clusters.  Using these results, we fit a scaling relation of the same form as equation~(\ref{eq:lambda-mass}) to the corrected data.  Table~\ref{tab:results} contains the best fit parameters and one sigma uncertainties of the $\lambda$-- mass relation with the correlated scatter correction.
The impact of the correlated scatter in $\lambda$ and $L_\mathrm{X}$ is smaller than the $1\sigma$ statistical parameter uncertainty for all three parameters.  Thus, for a sample the size of the current SPIDERS analysis, this effect can be safely ignored.

\begin{figure}
\centering { 
   \includegraphics[scale=0.55]{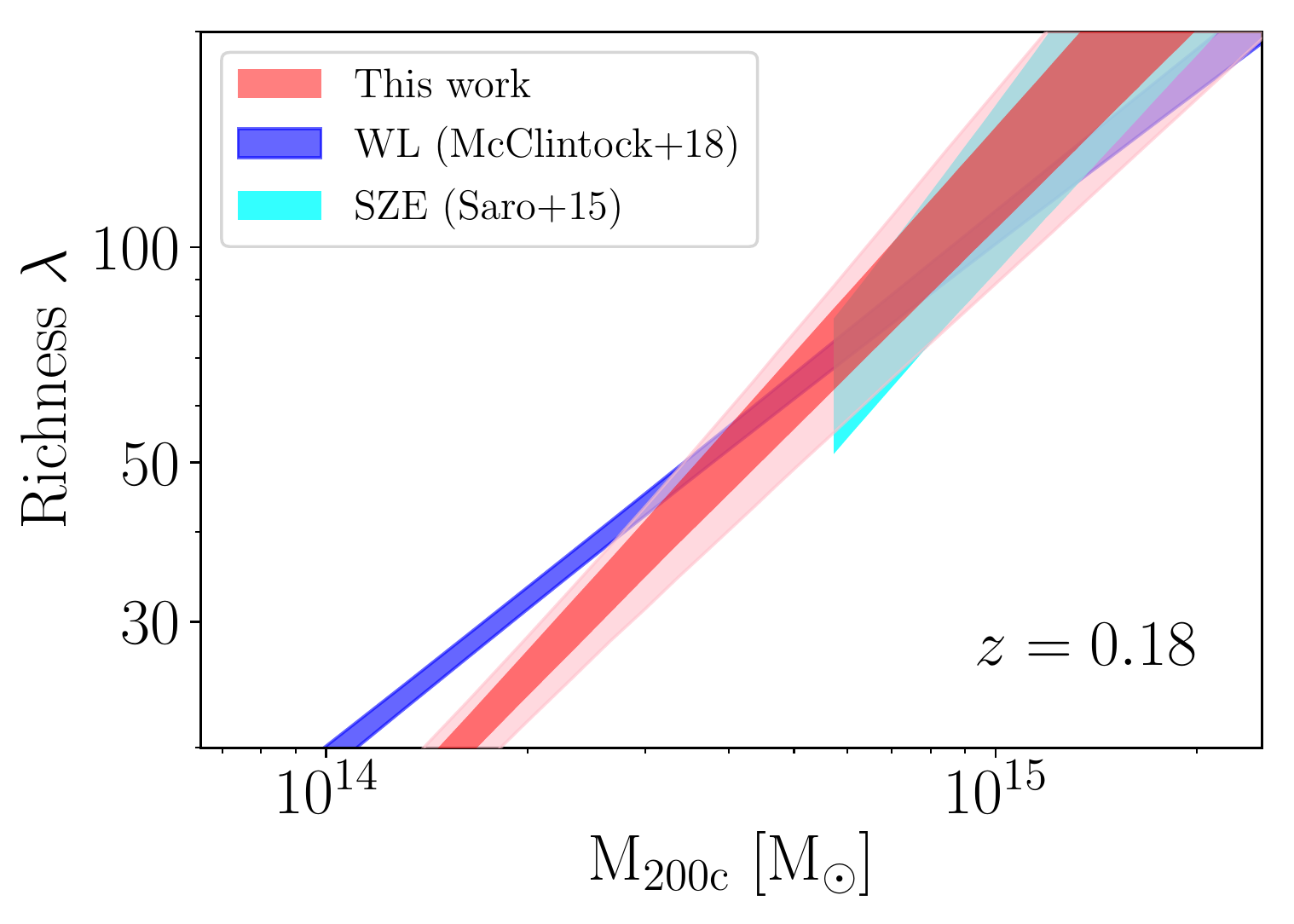} 
      }
\caption{Best fit model for our richness-mass relation (in red), evaluated at the redshift $z = 0.18$, compared to other measurements. For our analysis we also show the $2\sigma$ confidence area (pink region around the red relation). Confidence regions include statistical uncertainties only.}
\label{fig:rich_comp}
\end{figure}

\begin{figure}
\centering 
\includegraphics[scale=0.55]{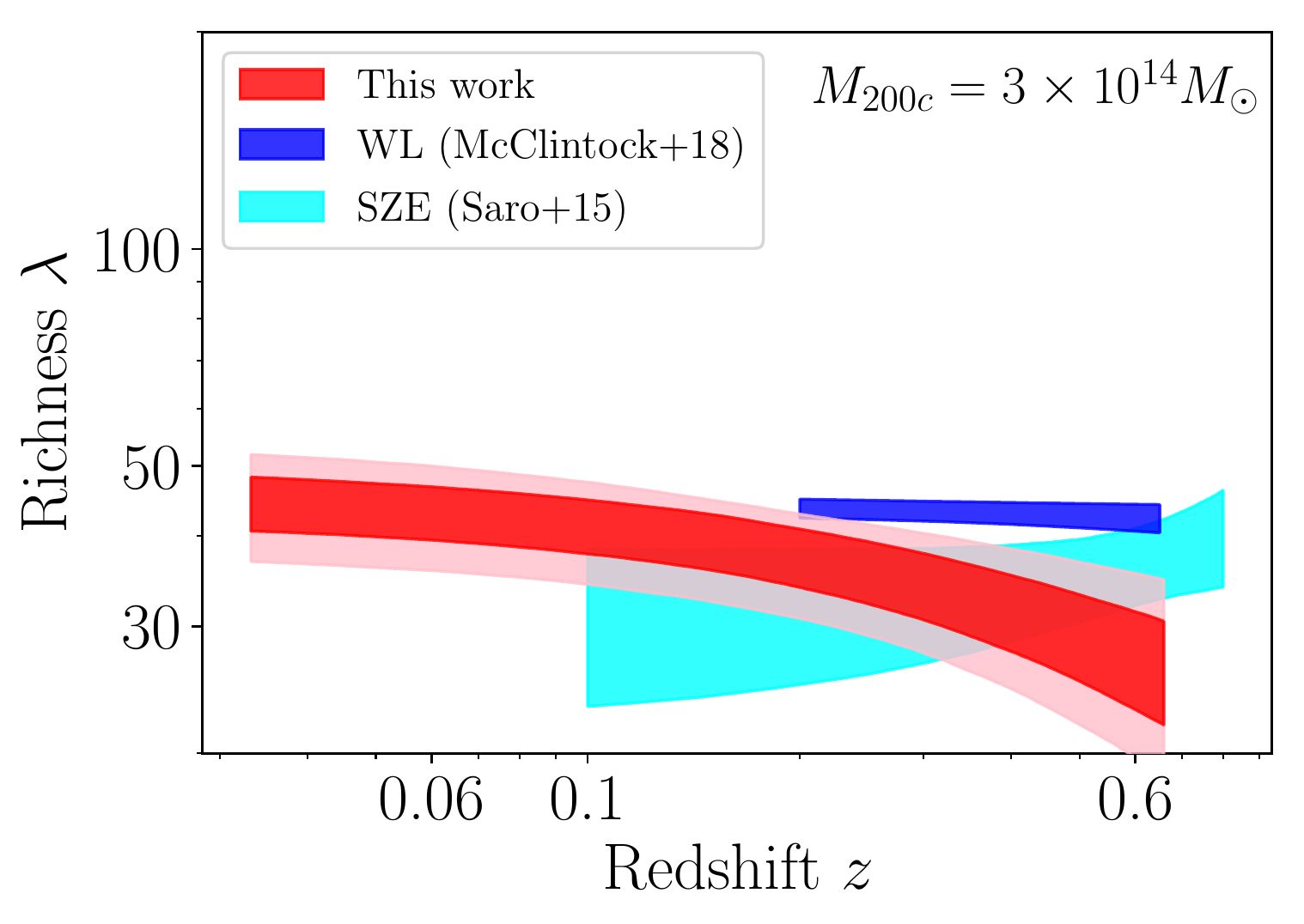}
\caption {Best fit model for our richness-redshift relation (in red), evaluated at our pivot mass 
$M_{\text{piv}} = 3 \times 10^{14} M_{\odot}$, compared to previous works. For our analysis we also show the $2\sigma$ confidence region. Confidence regions include statistical uncertainties only.}
\label{fig:z_evol}
\end{figure}


\subsection{Comparison to previous results}
\label{sec:comparison}

In this section we compare our calibration of the richness-mass relation to previous results from the literature. We show the mass and redshift trends of $\lambda$ in Figs.~\ref{fig:rich_comp} and~\ref{fig:z_evol}, respectively,  where for the redshift trend we correct the data points to the mass $M_\text{200c}=3\times10^{14}M_\odot$ and for the mass trend we move the data points to the redshift $z=0.18$.  These are the mass and redshift pivots of our sample, and are therefore the places where our constraints are tightest.
The best fit model for the $\lambda-M_{\text{200c}}$ relation is shown in red, with shaded $1\sigma$ and $2\sigma$ confidence regions. For the results from \citet[][in cyan]{2015Saro} and \citet[][in blue]{2018McClintock}, we show only the $1\sigma$ confidence region. We limit the redshift range to that analyzed in each work.  Fig.~\ref{fig:rich_comp} makes clear that the mass slope of our relation lies in between that of \citet{2015Saro} and \citet{2018McClintock}.  In Fig.~\ref{fig:z_evol} our results suggest stronger negative redshift evolution than in either previous results.

Table~\ref{tab:results} lists the parameter estimates and uncertainties for all the comparison results.  To make these comparisons, we scale all the measurements from previous analyses to the redshift $z_{\text{piv}} = 0.18$ (Fig.~\ref{fig:rich_comp}), and mass $M_{\text{piv}} = 3 \times 10^{14} M_{\odot}$ (Fig.~\ref{fig:z_evol}), using the best fit redshift and mass trends published for each sample.  Doing this, we predict the $\lambda(3 \times 10^{14} M_{\odot},0.18)$ for each previous work.   All mass conversions needed for the comparison plot are carried out using \textsc{Colossus}, an open-source python package for calculations related to cosmology \citep{2017Diemer}. The mass and redshift trend parameters presented in Table~\ref{tab:results} were also converted to those defined in equation~(\ref{eq:lambda-mass}) using the appropriate mass definition $M_\mathrm{200c}$ and redshift trend function $(1+z)^{\gamma_\lambda}$ adopted for our analysis here.  In some cases this involved inversions of the mass-observable relations.

Importantly, the definition of the cluster richness $\lambda$ from the redMaPPer algorithm may differ from one dataset to another.  Before comparing to our results, we implement this correction using the conversion obtained by \citet{2018McClintock}:
\be
\begin{split}
\lambda_{\rm{DES \, \, SV}} = & (1.08 \pm 0.16) \lambda_{\rm{DES \,\, Y1}}\\
\lambda_{\rm{SDSS}} = & (0.93 \pm 0.14) \lambda_{\rm{DES \,\, Y1}}
\end{split}
\ee
where the number presented as the uncertainty is actually the standard deviation in the richness ratio (thus, the uncertainty on the mean conversion factor is tiny in comparison).  We have applied these corrections to bring all results to the space of our analysis.

\subsubsection{Discussion of the mass trend parameter $\mathrm{B}_\lambda$}
\label{sec:Bdifference}

Our mass trend shows good agreement with the results obtained by \citet{2015Saro}, which is based on measurements of a cross-matched sample of SZE selected galaxy cluster candidates from the South Pole Telescope 2500~deg$^2$ SPT-SZ survey and the optically selected redMaPPer clusters from the Dark Energy Survey Science Verification (DES-SV) data. We also find good agreement with the scaling relation obtained by \citet{2016Baxter} and \citet{2018Baxter}, where the first is based on cluster clustering using SDSS data, and the second on Cosmic Microwave Background (CMB) lensing measurements from SPT in combination with DES Y1 redMaPPer clusters.  

On the other hand, our results are in disagreement with those of \citet{2017Simet}, based on redMaPPer clusters found in the Sloan Digital Sky Survey, and of \citet{2018McClintock}, obtained analyzing redMaPPer galaxy clusters identified in the Dark Energy Survey Year 1 data. While our analysis is performed on ensembles consisting of single clusters, these two analyses made use of stacked weak lensing data. In fact, neither of these analyses aimed to account for the Eddington bias and, therefore, they do not solve for the underlying richness-mass relation as we have done. Rather, they fit the mean $\lambda$ within bins of lambda and redshift to the mean weak lensing mass associated with each bin.  Because the Eddington bias is a function of the scatter in $\lambda$ and the effective slope of the mass function at the corresponding mass, ignoring the Eddington bias correction will lead to systematic errors in the redshift and mass trends.  We estimate that the Eddington bias correction will impact the mass and redshift trends with $\delta B_\lambda= +0.04$ and $\delta \gamma_\lambda=+0.09$, respectively, where $\delta$ is defined as the value of the parameter after applying the bias correction minus the one before the correction. 
With these corrections, the expected parameters for the underlying $\lambda$-mass 
relation would be $B_\lambda=0.77$ and $\gamma_\lambda=-0.01$.  These are still offset significantly from our measured values at $\Delta B_\lambda=-0.21\pm0.08$ (2.7$\sigma$) and $\Delta \gamma_\lambda=+1.12\pm0.60$ (1.9$\sigma$), and so clearly the Eddington bias is not large enough to explain the differences between the results.

We note that redMaPPer optical selection and RASS X-ray selection followed by cross-matching to redMaPPer (i.e., the CODEX sample we analyze here) will not generally lead to similar levels of sample contamination.  Moreover, contamination would be expected to have a different impact on a stacked weak lensing analysis than on a cluster by cluster dynamical analysis like that carried out here.  Thus, in principle, differences in the $\lambda$-mass relations constrained from these two different approaches can be used to shed light on the differences in contamination.  

The contamination of optically selected cluster samples by projected collections of passive galaxies in low mass groups and isolated systems has long been a concern \citep{2007Gladders,2012Song,2018Costanzi}, with estimates of contamination fractions reaching as high as $\sim$50\%.  Within X-ray imaging surveys like those employing pointed PSPC observations with $\sim25''$~FWHM imaging \citep[e.g.][]{1998Vikhlinin, 2018Clerc}, the selection of X-ray sources exhibiting extended emission has been shown to deliver contamination at the $\sim$10\% level.  Within the lower quality RASS imaging, where there is generally no extent information for the faint CODEX sources, the contamination is driven by random superpositions between the faint X-ray sources ($\sim90$\% are AGN or stars) and the ubiquitous red sequence optical candidate clusters identified by redMaPPer \citep[see detailed discussion of this problem and the description of a method to control this contamination in][]{2018Klein,2019Klein}.  

Within a stacked weak lensing analysis, these contaminating low mass systems would likely suppress the mass at a given $\lambda$, and a mass dependent contamination that increases toward low $\lambda$, as suggested by some studies \citep{2015Saro}, could lead to a significant bias to low values in the mass slope $\mathrm{B}_\lambda$.  Within this context, it is interesting to note that the disagreement in the $\lambda$-mass relations between \citet{2018McClintock} and our analysis is largest at low lambda.

For the CODEX sample, the random superpositions are not necessarily contaminants in a study of the $\lambda$--mass relation, because many of these random superpositions are of X-ray AGN projected to lie near true red sequence clusters on the sky.  Subsequent spectroscopic followup of these systems, whether the X-ray emission is AGN or cluster dominated, leads to dynamical sampling of clusters and groups, with less impact from the tail of low mass, contaminating structure projections than in the case of the purely optically selected sample.  Spectroscopic followup further reduces the contamination, because those systems that are loose projections can in many cases be separated out from the true, collapsed halos during the SPIDERS validation procedure \citep[see also detailed spectroscopic studies of redMaPPer systems in][]{2018Sohn,2018Rines}.

Because our dynamical analysis uses (weak) mass information from all individual systems, the impact of the final remaining contamination in the CODEX calibration of the $\lambda$-mass relation, which would tend to be sampled with smaller numbers of spectroscopic redshifts, would then be further reduced.  Thus, because both methods-- optical cluster selection + stacked weak lensing and RASS+optical redMaPPer + dynamics-- are subject to different systematic effects, we have a potential explanation for the different mass slopes observed in the two analyses.  Further work using structure formation simulations or generation of realistic mocks including the appropriate contamination effects would be required to quantify these effects and understand the differences in detail.  Supplementing this with dense spectroscopic studies of redMaPPer samples to better understand the nature of the projection and contamination issues will also be very helpful \citep{2018Sohn,2018Rines}.

Finally, we compare our scaling relation amplitudes and mass trends with those obtained by two recent low redshift ($z\leq0.33$) SDSS based analyses. 
\citet{2018Murata} perform a richness-mass scaling relation calibration using a joint measurement of the abundance and stacked cluster weak lensing profiles within the context of the cosmological parameters preferred by Planck CMB anisotropy \citep{2015Planck}. They determine a scaling relation that reproduces both the cluster counts and the lensing profiles but only at very large richness scatter $\sigma_{\ln{\lambda{\left|M\right.}}}=0.46$.  Scatter of this scale predicts a non-negligible contribution of low-mass haloes ($M_{\rm{200m}} \lesssim 10^{13} M_{\odot} $) in the SDSS redMaPPer sample. Their interpretation is that this contamination could be due to projection effects that preferentially impact the low richness portion of the sample ($20\le\lambda\le30$) or that the assumed Planck cosmology is different from the true underlying cosmology.  We find good agreement with the mass trend of their results, but their amplitude is only about half of the value we find.  The offsets in amplitude are not surprising given the very large differences in the scatter in the two analyses.

\citet{2017Jimeno} calibrate the mass-richness scaling relation using both the cluster correlation function and the cluster counts.  They employ the N-body Millennium XXL simulations, updated to the Planck cosmology \citep{2015Planck} to predict the distributions of clusters in richness.  They first obtain two independent mass-richness relations using separately clustering and counts data, and afterwards perform a joint analysis. Interestingly, they find a 2.5$\sigma$ tension between the amplitudes of the scaling relation in the two cases that weakens if they shift from the Planck cosmological parameters to those from the WMAP mission \citep{2003Spergel}.  The joint constraints on the amplitude and mass trend of the mass-richness relation are in good agreement with our results.

Overall, the agreement with the counts+clustering analysis is encouraging, suggesting that their modeling of the redMaPPer selection and contamination cannot be far off.   However, the counts+stacked weak lensing analysis seems to provide further indications that projection effects in the redMaPPer sample may be responsible for differences between stacked weak lensing constraints and measurements of the true underlying richness-mass relation from direct mass measurements (our analysis), from counts or from cluster clustering.  Commonalities between the impact of correlated large scale structure on weak lensing and richness measurements may lie at the heart of these differences.

\subsubsection{Discussion of the redshift trend parameter $\gamma_\lambda$}
\label{sec:gammadifference}

Our constraint on the redshift trend of the $\lambda-M_{\text{200c}}-z$ relation shows a stronger negative trend $\gamma_\lambda=-1.13\pm0.33\pm0.49$ than found in previous analyses (Fig.~\ref{fig:z_evol}), which have provided no significant evidence of a redshift trend \citep{2015Saro,2018McClintock}. The behavior we see in the CODEX sample would be expected if there were an increasing fraction of red sequence (RS) galaxies over cosmic time, with no evolution in the overall halo occupation number $N_{200}$ of galaxies within the virial region above a particular stellar mass or luminosity cut. The redshift trend we measure is in rough agreement with results from \citet{2017Hennig}, a study of the galaxy populations in 74 SPT clusters whose redshifts extend to  $z\sim1.1$ and that were imaged as part of the DES SV survey. 
They find that the number of red sequence galaxies $N_\mathrm{200,RS}$ brighter than $m_*+2$ and within $r_\mathrm{200c}$ decreases with redshift at fixed mass as $N_\mathrm{200,RS} \propto (1+z)^{-0.84\pm0.34}$, corresponding to an evolution of the red sequence fraction within $r_\mathrm{200c}$ going as $f_{\rm{RS}} \propto (1+z)^{-0.65\pm0.21}$.  This evolution is less steep than the $\lambda$-mass evolution we observe here, but the two results are statistically consistent with a difference of $0.48\pm0.63$.  

In contrast, the \citet{2018McClintock} and \citet{2015Saro} results show no redshift trend with $\gamma_\lambda=-0.22\pm0.22$ and $0.60\pm0.63$, respectively.  These results differ from our measurement at 1.5$\sigma$ ($0.91\pm0.63$) and 2$\sigma$ ($1.73\pm0.86$), respectively.  Interestingly, as discussed in Section~\ref{sec:memcut}, our measured redshift trend is closer to that measured in the other two analyses when we include more high redshift clusters that are sampled by smaller numbers of spectroscopic redshifts.  Clearly, further study is needed to better understand whether there is a difference in the redshift trend inferred from dynamical masses and to pinpoint any underlying causes.


\section{Conclusions}
\label{sec:conclusions}

In this paper, we present a richness-mass-redshift scaling relation calibration using galaxy dynamical information from a sample of 428  CODEX galaxy clusters. These are X-ray selected systems from RASS that have red sequence selected redMaPPer optical counterparts within a search radius of 3$'$.  
Our sample has redshifts up to $z\sim0.66$ and optical richnesses $\lambda \ge 20$. The spectroscopic follow-up comes from the SPectroscopic IDentification of eRosita Sources (SPIDERS) survey, resulting in 7807 red member galaxies after interloper rejection and the exclusion of all systems with fewer than 10 member redshifts.

We study the $\lambda$-$M_\mathrm{200c}-z$ relation by extracting the likelihood of consistency between the velocity sample for each individual cluster and the modeled projected phase space velocity distribution for a cluster of inferred mass $M_\text{200c}$ given its observed $\lambda$ and redshift $z$. The modeling is carried out using a Jeans analysis based on the code MAMPOSSt \citep{2013MAMPOSSt}, which allows us to build the projected phase space velocity distributions for clusters of particular mass, given a range of models for the orbital anisotropy of the galaxies. In our analysis, we adopt an NFW mass profile and employ five different velocity dispersion anisotropy profiles. Furthermore, we adopt an NFW profile for the red galaxy tracer population with concentration $c=5.37$ \citep[][and Section~\ref{sec:number_density}]{2017Hennig}.  We combine results from the different anisotropy models by performing Bayesian model averaging, allowing us to effectively marginalize over the orbital anisotropy of the spectroscopic galaxy population.
 
We model the scaling relation as $\lambda \propto \text{A}_{\lambda}  {M_{\text{200c}}}^{\text{B}_{\lambda}}({1+z})^{\gamma_{\lambda}}$ (equation~\ref{eq:lambda-mass}). As described in Section~\ref{sec:scalingrelation}, we apply corrections for the Eddington bias and for the Malmquist bias.
Results are presented in Table~\ref{tab:results}.  For clusters at our pivot redshift of $z_{\text{piv}} = 0.18$ and pivot mass of $M_{\text{piv}} = 3 \times 10^{14} M_{\odot}$, we find our constraints on the scaling relation to be as follows: the normalization $\text{A}_\lambda$, mass slope $\mathrm{B}_{\lambda}$ and redshift slope $\gamma_{\lambda}$ are 
\be 
\begin{split}
\mathrm{A}_{\lambda}= & 38.6^{+3.1}_{-4.1}\pm3.9,\\
\mathrm{B}_{\lambda}= & 0.99^{+0.06}_{-0.07}\pm0.04,\\
\gamma_{\lambda}= & -1.13^{+0.32}_{-0.34}\pm0.49. 
\end{split}
\ee
As discussed in Section~\ref{sec:systematics}, the quoted uncertainties include a 10\% 
systematic uncertainty on the dynamical mass that is applied wholely to the scaling relation amplitude \citep[see study of systematics in][]{2013MAMPOSSt} and a systematic uncertainty of 0.49 on the redshift trend $\gamma_\lambda$, that arises from sensitivity in our redshift trend parameter to cuts on the cluster sample according to the number of member galaxies with spectroscopic redshifts. 

Our results on the mass trend of the $\lambda$-mass scaling relation are in generally good agreement with previous studies of the mass dependence of the halo occupation number, or the number of cluster galaxies within a common portion of the luminosity function (often $m_*+2$) and within a common portion of the cluster virial region (typically defined using $r_{500}$ or $r_{200}$) \citep{2004Lin,2017Hennig}.
This is an indication that the redMaPPer algorithm is effective at selecting cluster galaxies over a common portion of the virial region and that the galaxy red sequence fraction is not a strong function of cluster mass in this mass range.

Moreover, our results are in good agreement with those from previous studies of the $\lambda$-mass relation using SPT selected clusters that have been cross-matched with DES SV identified optical systems \citep{2015Saro}. We are also consistent with the value of the mass trend measured using cluster clustering in SDSS \citep{2016Baxter} and CMB lensing of the DES Yr 1 redMaPPer sample using SPT \citep{2018Baxter}.  On the other hand, our results are in disagreement with a study of
redMaPPer clusters detected in SDSS data \citep{2017Simet} and show a $\sim 2.7\sigma$ 
tension with the constraints obtained from redMaPPer galaxy clusters identified in the DES Y1 data \citep{2018McClintock}.  
Both of these latter results arose from the analysis of stacked weak lensing signatures, and neither analysis sought to obtain the true underlying $\lambda$-mass relation after correction for the Eddington bias.  As discussed in Section~\ref{sec:Bdifference}, the Eddington bias correction would not be large enough to explain the difference.  We suggest instead that the difference is reflective of the likely differences in the contamination of a pure redMaPPer sample and our CODEX sample, which is first X-ray selected and then cross-matched to the redMaPPer candidates within 3$'$ radius.

In Section~\ref{sec:Bdifference} we also discuss two scaling relation calibrations that adopt redMaPPer counts together with either stacked weak lensing  or cluster clustering to calibration the richness-mass relation.  Inferring cluster mass information from the counts such as in those two analyses requires an accurate description of the contamination or projection effects in the redMaPPer sample.  Interestingly, our dynamical mass calibration results are in good agreement with the counts+clustering analysis \citep{2017Jimeno}, but not with the counts+stacked weak lensing analysis \citep{2018Murata}, where the authors find a dramatically larger scatter in richness-mass is required to bring their weak lensing and counts constraints on cluster masses into agreement.

The redshift trend $\gamma_{\lambda}$ of our richness-mass relation shows a strong 
negative trend where $\lambda$ at fixed mass decreases with redshift. This result can 
be interpreted as an indication of the increasing fraction of cluster red sequence galaxies over cosmic time.  As presented in Section~\ref{sec:gammadifference}, our results are somewhat steeper than but statistically consistent with those from \citet[][]{2017Hennig}, where they studied SPT selected clusters and found that the fraction of red sequence galaxies to $m_*(z)+2$ decreases with redshift, from $\sim 80\%$ at $z \sim 0.1$ to $\sim 55\%$ at $z \sim 1$, following the form $f_\mathrm{RS} \propto (1+z)^{-0.65\pm0.21}$.   
Our measurement is steeper than other results showing little or no redshift trend in the $\lambda$-mass relation \citep{2015Saro,2018McClintock}, but the differences are only significant at 1.5 and 2$\sigma$, respectively.  Further study of the redshift trend of the $\lambda$--mass relation is clearly warranted.

In addition, we test the impact of interesting selection effects on our results in Section~\ref{sec:correlatedscatter}.  We show that negative covariance between the scatter  in X-ray luminosity and the scatter in optical richness for clusters at the levels measured in  the CODEX sample has negligible impact on the $\lambda$-mass relation.  

In summary, dynamical masses are a powerful tool to gain information on the link 
between the masses of galaxy clusters and readily obtainable observables-- even 
in the limit of large cluster samples with small spectroscopic samples available for 
each cluster.  Further work to perform a dynamical analysis on numerical simulations 
of structure formation will be crucial to being able to properly assess the 
true precision and robustness of the dynamical masses and anisotropy 
measurements we seek to extract from the data.  A better understanding of the expected variation of the velocity anisotropy profile, of the distribution of interlopers after cleaning and of the impact of departures from equilibrium on our Jeans analysis will be broadly helpful.  Our analysis demonstrates that there is promise in the analysis of small per-cluster spectroscopic samples of the sort that will be delivered by future spectroscopic surveys like DESI \citep{levi13}, 4MOST \citep{dejong12} and Euclid \citep{laureijs11}.


\section*{ACKNOWLEDGMENTS}
RC would like to thank Gus Evrard, Arya Farahi, Tom McClintock and Steffen Hagstotz for helpful discussions.  The Munich group acknowledges the support by the DFG Cluster of Excellence ``Origin and Structure of the Universe'', the Transregio program TR33 ``The Dark Universe'', the MPG faculty fellowship program and the Ludwig-Maximilians-Universit\"at Munich. RC acknowledges participation in the IMPRS on Astrophysics at the Ludwig-Maximilians University and the associated financial support from the Max-Planck Society. RC and VS acknowledge support from the German Space Agency (DLR) through \textit{Verbundforschung} project ID 50OR1603. AS is supported by the ERC-StG ``ClustersXCosmo", grant agreement 716762.   AB acknowledges the hospitality of the LMU and partial financial support from PRIN-INAF 2014 ``Glittering kaleidoscopes in the sky: the multifaceted nature and role of Galaxy Clusters?", P.I.: Mario Nonino. 

Funding for the Sloan Digital Sky Survey IV has been provided by the Alfred P. Sloan Foundation, the U.S. Department of Energy Office of Science, and the Participating Institutions. SDSS-IV acknowledges support and resources from the Center for High-Performance Computing at the University of Utah. The SDSS web site is www.sdss.org.

SDSS-IV is managed by the Astrophysical Research Consortium for the Participating Institutions of the SDSS Collaboration including the Brazilian Participation Group, the Carnegie Institution for Science, 
Carnegie Mellon University, the Chilean Participation Group, the French Participation Group, Harvard-Smithsonian Center for Astrophysics, Instituto de Astrof\'isica de Canarias, The Johns Hopkins University, Kavli Institute for the Physics and Mathematics of the Universe (IPMU) / 
University of Tokyo, the Korean Participation Group, Lawrence Berkeley National Laboratory, 
Leibniz Institut f\"ur Astrophysik Potsdam (AIP),  
Max-Planck-Institut f\"ur Astronomie (MPIA Heidelberg), 
Max-Planck-Institut f\"ur Astrophysik (MPA Garching), 
Max-Planck-Institut f\"ur Extraterrestrische Physik (MPE), 
National Astronomical Observatories of China, New Mexico State University, 
New York University, University of Notre Dame, 
Observat\'ario Nacional / MCTI, The Ohio State University, 
Pennsylvania State University, Shanghai Astronomical Observatory, 
United Kingdom Participation Group,
Universidad Nacional Aut\'onoma de M\'exico, University of Arizona, 
University of Colorado Boulder, University of Oxford, University of Portsmouth, 
University of Utah, University of Virginia, University of Washington, University of Wisconsin, 
Vanderbilt University, and Yale University.

\bibliographystyle{mn2e}
\bibliography{literature}

\end{document}